\documentclass{iopjournal}

\usepackage{fancyhdr}
\usepackage{amsmath}
\usepackage{amssymb}
\usepackage{graphicx}% Include figure files
\usepackage{dcolumn}% Align table columns on decimal point
\usepackage{bm}% bold math
\usepackage{xcolor}
\usepackage{hyperref}% add hypertext capabilities
\usepackage{orcidlink}
\usepackage{cite}

\newcommand{\new}[1]{#1}

\newcommand{\dt}[1]{\frac{\text{d}#1}{\text{d}t}}

\begin{document}

\articletype{Paper} %	 e.g. Paper, Letter, Topical Review...

% earlier suggestion: Stochastic Activation in Vesicle-Mediated Signaling Shaped by Arrival Statistics
\title{Activation in Vesicle-Mediated Signaling Shaped by Batch Arrival Statistics}

\author{Jan Hauke$^{1,\dagger}$\orcidlink{0009-0003-6032-9852}
, Julian B. Voits$^{1,2,\dagger}$\orcidlink{0009-0006-2650-6968} and Ulrich S. Schwarz\textsuperscript{1,2,*}\orcidlink{0000-0003-1483-640X}}

\affil{$^1$Institute for Theoretical Physics, Heidelberg University, Germany}

\affil{$^2$BioQuant-Center for Quantitative Biology, Heidelberg University, Germany}

\affil{$^\dagger${These authors contributed equally to this work.}}

\affil{$^*$Author to whom any correspondence should be addressed.}

\email{schwarz@thphys.uni-heidelberg.de}


\begin{abstract}
Vesicle-mediated secretion of ions or molecules is a central mechanism of cellular communication, for example
in processes such as neurotransmission or hormone release. 
These events are inherently stochastic: vesicle fusions lead to bursts of variable sizes, releasing discrete packets of transmitters that are subsequently cleared or degraded. The dynamics \new{are intrinsically time-directed} due to the interplay of spontaneous bursts and continuous degradation. 
Using generating functions and a recursion relation, we derive
an exact solution for the full time-dependent probability distribution 
of a general batch arrival-degradation model. 
This framework also enables a full analysis of 
first-passage times to a concentration threshold representing downstream activation.
We show that activation kinetics are not determined by mean dynamics alone, but depend sensitively on the temporal statistics of arrival events, batch-size variability, and degradation. In particular, different arrival processes with identical mean rates can lead to qualitatively distinct first-passage behavior, reflecting the role of time-asymmetric fluctuations. 
We also discuss extensions incorporating vesicle depletion. 
Our results provide a transparent link between stochastic release dynamics and 
activation timing in vesicle-mediated signaling.
\end{abstract}

\section{\label{sec:level1}Introduction}

\begin{figure*}[t]
    \centering
    \includegraphics[width=\textwidth]{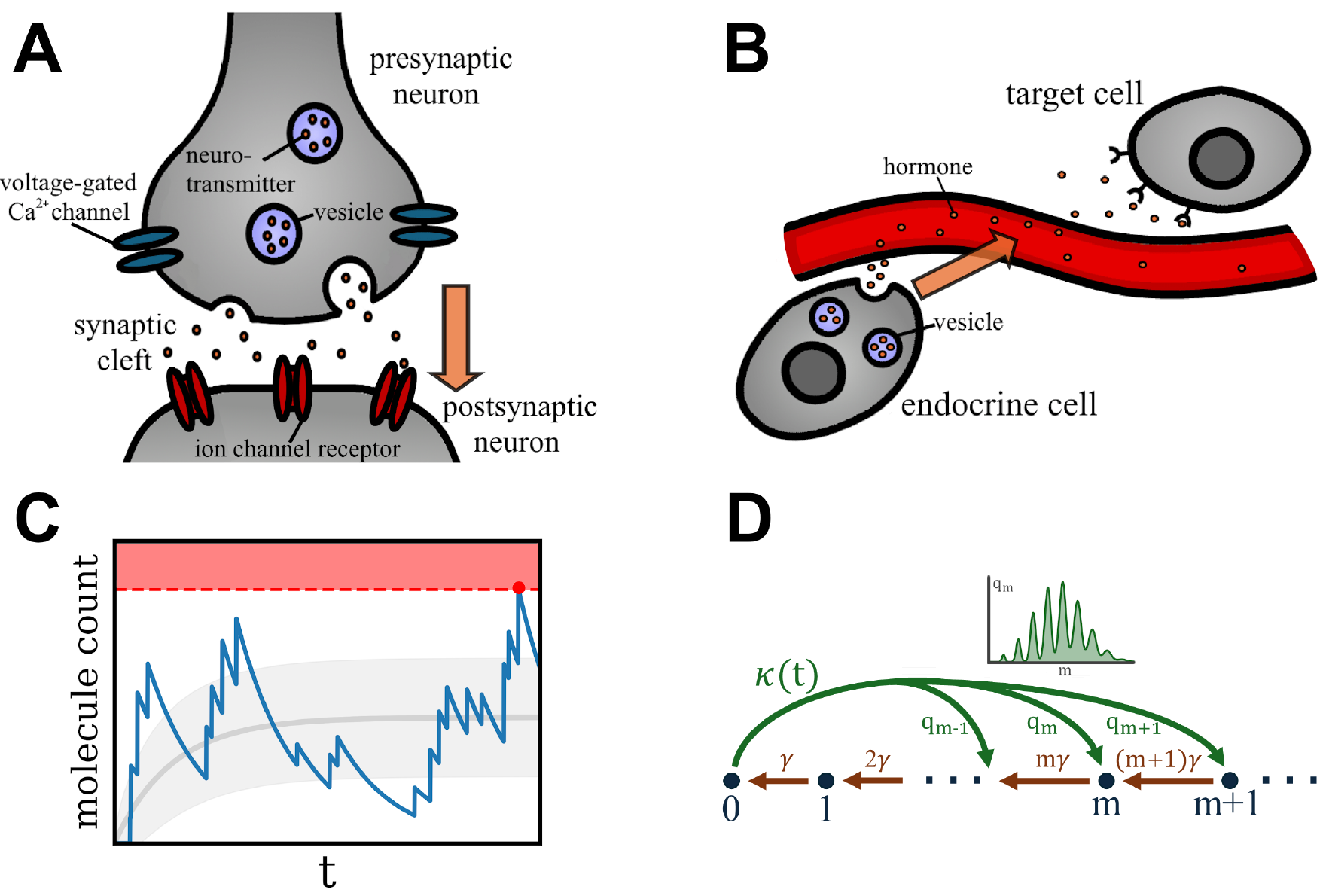}
    \caption{Stochastic burst–degradation dynamics in cellular communication.
    \textbf{A} Synaptic signaling: Neurotransmitter molecules are released in discrete batches from synaptic vesicles into the synaptic cleft following neuronal stimulation. Molecules diffuse, bind to postsynaptic receptors, and are cleared by uptake or enzymatic degradation.
    \textbf{B} Endocrine signaling: Hormone-containing vesicles in an endocrine cell release bursts of molecules into the bloodstream. The hormones are transported, degraded, or diluted, and activate distant target cells once the local concentration exceeds an effective threshold. \textbf{C} The molecule count in these systems becomes a stochastic variable (blue line) that is shaped by batch-release jumps and exponential decay due to degradation. Such a stochastic description is crucial to describe downstream activation that is triggered if the molecule count exceeds a certain value (red area), as deterministic predictions and typical variations around it (gray area) may fail to capture the threshold-crossing event. \textbf{D} Schematic of 
    the stochastic burst–degradation model. Molecules are removed from the system at a degradation rate $\gamma$, while bursts occur at a (generally time-dependent) rate $\kappa(t)$. Each burst adds $m$ molecules to the system with probability $q_m$, which tends to be a multimodal distribution as a consequence of multi-vesicle release. The number of released molecules $m$ from a burst is added to the number $n$ of existing molecules.}
    \label{fig:biological_motivation}
\end{figure*}

\new{Biological cells communicate through many
different channels, including biochemical, 
adhesive and mechanical signals \cite{alberts2022molecular}. The prototypical
type of cell-cell communication is the release of a
biochemical signal, e.g. a growth factor
or a cytokine, into the surrounding
medium, which is then sensed and processed
by other cells \cite{su2024cell}. In some biological situations
of interest, the release process occurs
through vesicle secretion, such that large
numbers of molecules are released at the same
time. One example is the neuronal synapse,
where signaling is very local, from the
presynaptic to the postsynaptic cell, as shown
schematically in Fig.~\ref{fig:biological_motivation}A \cite{sudhof2004synaptic,sudhof2011synaptic, rizo2015synaptic}.
Another example is endocrine signaling,
where hormones are released into the bloodstream
for long-range signaling in the body, as shown 
schematically in Fig.~\ref{fig:biological_motivation}B \cite{lang1999molecular}.}
In either case, signaling molecules are released through stochastic vesicle fusion events, leading to a discrete, quantized increase in the local molecular count, and are removed through transport and degradation processes \cite{bergles1999clearance}. The resulting dynamics are therefore shaped by spontaneous increases (bursts) rather than showing smooth deterministic trajectories \cite{brabant1992pulsatile,veldhuis2008motivations, ernst2022variance,ernst2023rate}. Postsynaptic neurons or hormone-receiving cells respond once the molecular count exceeds a certain threshold, making the timing of such threshold crossings a central observable \cite{johnston1994foundations,lodish2008molecular,ghusinga2015theoretical,gerstner2014neuronal} \new{(Fig.~\ref{fig:biological_motivation}C)}. From a physical perspective, vesicle-mediated signaling is an example of a strongly non-equilibrium transport process, where the molecular count is shaped by spontaneous burst events and continuous degradation. \new{This leads to intrinsically 
time-directed dynamics,
which are not described well by mean production and degradation rates alone.}

\new{The problem of the molecule count reaching a critical threshold can be naturally formulated in terms of first-passage times (FPT) describing the stochastic time required to reach a target for the first time \cite{redner2001guide, chou2014first, bressloff2014stochastic}. While FPTs are relatively well studied for diffusion processes and continuous stochastic dynamics \cite{metzler2014first, benichou2014first}, the burst-driven systems require a discrete description shaped by the interplay of spontaneous releases, burst size statistics and molecular degradation.
Recent work has made significant progress in understanding these contributions, including a derivation of exact first-passage time distributions in stochastic protein accumulation models \cite{rijal2022exact} and analytical descriptions of quantal release statistics in synaptic transmission \cite{rijal2024exact}. These studies highlighted the importance of stochasticity in such burst-degradation systems, but primarily focused on steady-state or time-homogeneous settings. However, a general model that incorporates both arbitrary release statistics and time-dependent rates to describe stochastic molecule counts and threshold-crossing events has
not been analyzed yet.}

\new{To incorporate these aspects into a general framework, we here study a general batch arrival–degradation process as a minimal mathematical model for such systems  \cite{gillespie2005evolution,daw2019distributions,bhattacharya2021random}. The molecule count is described as a stochastic variable, which is increased in discrete jumps due to spontaneous release events and decreased by ongoing degradation (Fig.~\ref{fig:biological_motivation}D). Such models also occur in bursty gene expression, where bursts arise from intermittent promoter activity and finite mRNA lifetimes. In fact, there is a significant body of experimental \cite{elowitz2002stochastic, ozbudak2002regulation, raser2004control,raser2005noise,golding2005real,cai2006stochastic,raj2008nature} and theoretical \cite{tunnacliffe2020transcriptional, cao2020analytical, rijal2022exact, ham2024stochastic, szavits2024solving, wang2025joint} literature on bursts in gene expression, often also addressing the relation between microscopic production kinetics and macroscopic variability and activation timing.
However, in marked contrast to vesicle-mediated release, transcriptional or translational burst sizes of gene expression are geometrically distributed, often with time-homogeneous production and degradation rates~\cite{paulsson2005models,shahrezaei2008analytical}, while the molecular count of a single vesicle} is more naturally described by binomial statistics, reflecting the finite number of release sites and probabilistic vesicle fusion \cite{del1954quantal,katz1969release, veldhuis2008motivations, hatamie2024insulin, malagon2016counting}.

\new{The framework analyzed here allows us to consider arbitrary burst profiles, including binomial distributions describing single-vesicle releases, but also more} complex and potentially multimodal distributions, \new{which} can arise due to multivesicular release and heterogeneity in release probability or vesicle content \cite{silver2003estimation, rudolph2015ubiquitous}. Moreover, for neurons in particular, the vesicle release rate is generally time- and history-dependent, reflecting short-term memory driven by presynaptic $\text{Ca}^{2+}$ dynamics and vesicle-pool depletion \cite{zucker2002short,regehr2012short}.
\new{In an extension of our model, we show that vesicle} depletion can be included in the description by treating the number of docked vesicles \new{(which are the vesicles that are
already positioned at the membrane and which can readily fuse upon the arrival of an action potential)} as an additional stochastic variable, typically with a constant replenishment rate \cite{pulido2015vesicular, rosenbaum2012short, loebel2009multiquantal, fuhrmann2002coding, rijal2024exact,ali2025exact, gambrell2024feedforward, gambrell2025analysis}, although extensions to include molecule-dependent rates, vesicle undocking and repair steps have also been proposed \cite{gambrell2025modulation,zhang2015improved,zhang2020analysis, vahdat2025inferring,gambrell2025consequences}.  

\new{A central question that our model allows us to address is how different batch arrival statistics lead to qualitatively different downstream activation profiles as a consequence of non-trivial threshold-crossing behavior. In particular, we investigate how different time statistics of bursts shape activation for processes with identical mean behavior but different arrival statistics. We find that these differences alone can lead to qualitatively distinct activation profiles, highlighting the role of the batch-arrival statistics for first-passage times in these systems. Hence,} the resulting activation cannot be predicted from deterministic \new{dynamics} alone and requires a careful mathematical treatment of the resulting first-passage time distributions. 
\new{These findings are complementary to recent work which} has highlighted that intrinsic noise can be decisive in predicting the timing of cellular events \cite{ghusinga2017first, rijal2022exact}. In particular, Ham et al. \cite{ham2024stochastic} demonstrate that stochasticity can accelerate or delay the mean first-passage time relative to deterministic predictions. 

Our paper is organized as follows. In Sec.~\ref{subsec:model}, we introduce our model. 
In Sec.~\ref{sec:distribution}, we apply the generating function method to derive a general analytical solution for the time-dependent probability distribution, in terms of a recurrence relation between the occupation probabilities.
In Sec.~\ref{sec:FPTs}, we then express conditional threshold-crossing probabilities as a convolution of stochastic decay and the burst size distribution and specify the latter in terms of vesicle release statistics and quantal variability.
We then apply this framework to several biologically relevant scenarios: we compare fixed-interval and homogeneous Poisson spike trains and cover the example of an exponentially decaying release rate and show that variability in the timing of release events strongly facilitates threshold crossing. Moreover, we illustrate that stochastic trajectories can reach thresholds significantly earlier than suggested by their deterministic dynamics. Lastly, Sec.~\ref{sec:vesicle-depletion} extends the description to incorporate the effect of vesicle depletion and captures its impact as a correction to the release statistics and activation dynamics.
We finally close with a summary and discussion in Sec.~\ref{summary}.

\section{\label{subsec:model} Model}

We study a model in which vesicle release is assumed to occur independently at a (potentially time-dependent) rate $\kappa(t)$, corresponding to a general time-inhomogeneous Poisson process. Each vesicle fusion event triggers the release of $m$ molecules with probability $q_m$. \new{In interneuronal signal transmission, the triggering event for vesicle releases is the arrival of an action potential (which is a traveling wave of a transient reversal in the membrane potential) in the presynaptic neuron. This} can release multiple vesicles, so \new{the release size distribution} $q$ depends both on the number of vesicles released and the number of molecules per vesicle. The molecules are subsequently removed by dilution, absorption or degradation, occurring at a rate $\gamma$ per molecule, so that the total loss rate is proportional to the molecular count. This leads to the reaction network for a burst shown in Fig.~\ref{fig:biological_motivation}D. Note that
each burst adds $m$ molecules to the existing 
number $n$ of molecules.
The resulting dynamics are intrinsically asymmetric: the increase of the molecular count is driven by spontaneous bursts while it decreases as a result of ongoing degradation, leading to \new{asymmetric} stochastic dynamics with \new{unidirectional transitions}.

The probability $p_n(t)$ of having $n$ molecules in the environment at time $t$ obeys the following master equation:
\begin{align}
\label{eq:master_eq}
    \dt{p_n}=-(\kappa+\gamma n)p_n+\gamma (n+1)p_{n+1}+\kappa (q \ast p)[n]\ .
\end{align}
In the last term, we use the discrete convolution
to represent the different burst sizes:
\begin{align}
    (q \ast p)[n]=\sum_{m=0}^n p_{n-m}q_{m}=\sum_{m=0}^\infty p_{n-m}q_{m}\ .
\end{align}
Note that the sum can be extended 
because $p_n(t)=0$ if $n<0$.
While the two terms in Eq.~(\ref{eq:master_eq})
with $\gamma$ correspond
to a simple one-step process of continuous degradation,
the two terms with $\kappa$ correspond to
the bursts and balance each other, because
each burst takes from the state $n$ and 
gives to the state $n+m$. 

One can non-dimensionalize time by rescaling it as $\gamma t \rightarrow t$, introducing the dimensionless parameter 
$\lambda(t) := \frac{\kappa(t)}{\gamma}$, to obtain:
\begin{align}
  \dt{p_n}=-( n+\lambda)p_n+ (n+1)p_{n+1}+ \lambda(q \ast p)[n],\label{eq:master_equation}
\end{align}
leaving $\lambda(t)$ as the only free parameter of the system. 
Multiplying by $n$ and summing over $n$ on both sides of the master equation yields a differential equation for the mean molecule count:
\begin{align}
    \frac{d}{dt}\langle n \rangle = -\langle n \rangle + \lambda(t) \langle m\rangle,
\end{align}
where $\langle m\rangle =\sum_{m=0}^\infty mq_m$ denotes the average forward jump size. This is solved by
\begin{align}
    \langle n\rangle(t)= n_0e^{-t}+\langle m\rangle\int_0^t\text{d}s\;\lambda(s)e^{s-t}. \label{eq:me-mean-evolution}
\end{align}
For constant $\lambda$, this becomes
\begin{align}
    \langle n\rangle(t)= n_0e^{-t}+\langle m\rangle
    \lambda (1-e^{-t})
\end{align}
and the deterministic long-time limit is 
$\langle m\rangle \lambda$. 
However, as illustrated in Fig.~\ref{fig:biological_motivation}C, stochastic trajectories fluctuate around this mean and can transiently reach molecule counts significantly above the deterministic limit. 
This implies that random activation of downstream processes with higher thresholds becomes possible, highlighting the need for a full stochastic description to capture the system’s behavior accurately.

\new{In the general framework as formulated here, one can
choose any release size distributions $q$ of interest. 
We briefly discuss some common modeling choices. Following the paradigmatic example of vesicle release in a neuronal synapse, we decompose $q$ into the distribution $q^{(V)}$ of the count of released vesicles (\textit{quantal count}) and the distribution $q^{(C)}$ of the molecular vesicle content (\textit{quantal size}). Given that $v$ vesicles are released, the probability that $m$ molecules are released is $\left(q^{(C)}\right)^{*v}_m $, where $(q^{(C)} )^{*v}$ is the $v$-fold discrete convolution of the quantal size distribution $q^{(C)}$. Taking the expectation over the quantal count distribution $q^{(V)}$, one obtains 
\begin{equation}
    q_m =\sum_{v=0}^{\infty} \left(q^{(C)}\right)^{*v}_m \left(q^{(V)}\right)_v. \label{eq:release-size-distribution-def}
\end{equation}}
As can be easily verified, the mean release size $\langle m\rangle$ is the product of the mean quantal count $\langle v \rangle$ and the mean quantal size $\langle c \rangle$.

The widely adopted binomial model \cite{katz1969release, quastel1997binomial} assumes that upon the arrival of an action potential, vesicles fuse independently of each other, resulting in a binomial distribution $q^{(V)}_v= \text{Binom}(v \,|\, V,\rho)$, where $V$ is the number of docked vesicles and $\rho$ \new{is the individual probability for a docked vesicle to be released}. For now, we assume $V$ to be constant. An extension to settings where the rate of vesicle replenishment is on the order of the release rate, meaning that $V$ itself has to be interpreted as a stochastic variable \cite{hennig2013theoretical}, is given in Section~\ref{sec:vesicle-depletion}.

\begin{figure}[htbp]
  \centering
  \includegraphics[width=0.8\linewidth]{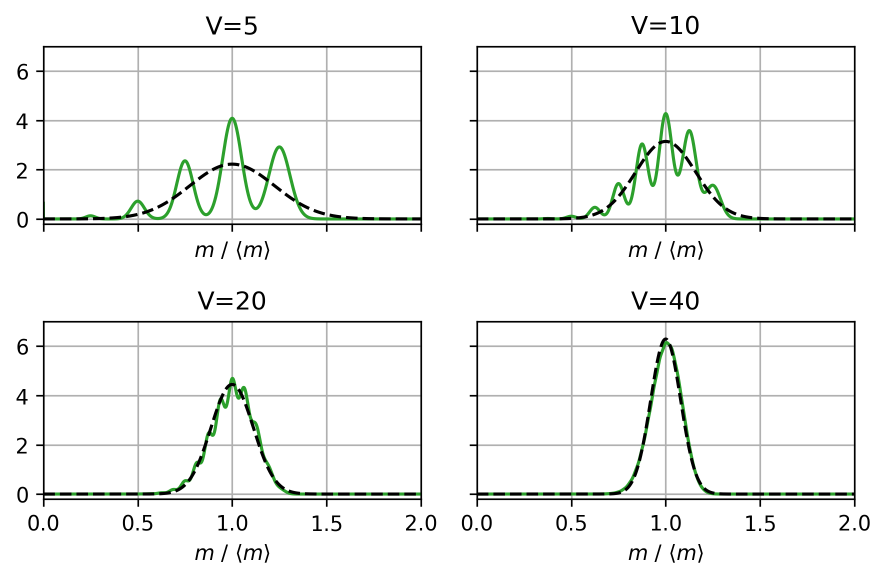}
  \caption{\new{Release size distribution $q$ (see Eq.~(\ref{eq:release-size-distribution-def})) with the number of vesicles distributed according to  $q^{(V)}= \text{Binom}( V,\rho\!=\!0.8)$ and the number of molecules per vesicle distributed according to $q^{(C)} = \mathcal{N}(\langle c \rangle,\sigma_c^2=(0.1\langle c \rangle)^2)$. The dashed line represents a normal distribution with mean $\langle m \rangle = \langle c \rangle V \rho $ and variance $\sigma_m^2$ as in Eq.~(\ref{eq:release-size-cov}).}}
  \label{fig:release-size}
\end{figure}

It is important to note that the quantal size is subject to fluctuations \cite{edwards2007neurotransmitter}, and its exact distribution is experimentally only partially accessible \cite{gordleeva2023estimation}.  \new{The variance $\sigma_c^2$ of the quantal size modifies the variance $\sigma_m^2$ of the total release size according to}
\begin{equation}
    \frac{\sigma^2_m}{ \langle m \rangle^2} = \frac{\sigma_v^2}{\langle v\rangle^2}+ \frac{1}{\langle v \rangle}\frac{\sigma_c^2}{\langle c \rangle^2}.
\end{equation}
For a binomial \new{quantal count distribution $q^{(V)}$}, this specializes to:
\begin{equation}
    \frac{\sigma^2_m}{ \langle m \rangle^2} = \frac{1}{V \rho} \left(1-\rho + \frac{\sigma_c^2}{\langle c \rangle^2} \right) \label{eq:release-size-cov}
\end{equation}
According to the binomial limit theorem, if \new{the number of releasable vesicles $V$ is not too small}, one may approximate $q$ by a normal distribution with mean $\langle m \rangle$ and variance $\sigma_m^2$ (see Fig.~\ref{fig:release-size}).

\section{\label{sec:distribution} Probability distribution of the molecule count}
The generating function of the probability distribution is defined as $G(z,t):=\sum_{n=0}^\infty p_n(t)z^n$  \cite{van1992stochastic}. Applying this transformation to Eq.~(\ref{eq:master_equation}) converts the master equation into a partial differential equation for $G(z,t)$:
\begin{align}
    \partial_t\ln{G}&=(1-z)\partial _z \ln{G}+\lambda(t)\big(\sum_{m=0}^\infty q_mz^m-1\big)\label{eq:PDE_generating_function}.
\end{align} 

\begin{figure}[t]
  \centering
  \includegraphics[width=0.8\linewidth]{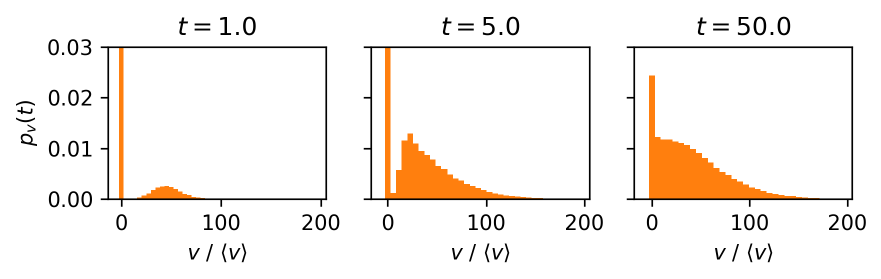}
  \caption{\new{Simulated time evolution of the molecule count distribution $p(t)$ for a constant release rate $\lambda(t) = 1$. Molecule counts $n$ are normalized by a fixed vesicle content of 1000 molecules, meaning $v := \frac{n}{1000}$, corresponding to a quantal size distribution of $q^{(C)}_m=\delta_{m,1000}$. The number of vesicles released per event follows $q^{(V)} = \text{Binom}(V=50, \rho=0.2)$, corresponding to $V=50$ docked vesicles each released independently with probability $\rho =0.2$, yielding a mean count of $\langle v \rangle = 10$.} One sees that the system first grows in bursts due to the initial condition of no molecules, and then relaxes again.}
  \label{fig:histograms}
\end{figure}

The solution for $G(z,t)$ subject to the boundary conditions $G(z=1,t)=1$ (for normalization) and for no initial molecular count ($p_n(t=0)=\delta_{n,0}$) is given by:
\begin{align}
    \ln G(z,t) &=\sum_{k=0}^\infty\sum_{m=k}^\infty q_m\big(\lambda\ast \mathcal{B}_k^m\big)(t)z^k-\int_0^t\text{d}s~\lambda(s), \label{eq:lnG-solution}
\end{align} 
where $\ast$ denotes the convolution in time and
\begin{align}
    \mathcal{B}_k^m(t):=\binom{m}{k}e^{-kt}(1-e^{-t})^{m-k}
\end{align}
is the binomial probability that $k$ out of $m$ molecules remain after time $t$. See Supplement IA for details on the derivation.
For an arbitrary initial condition $p_n(t=0)=\delta_{n,n_0}$, this result generalizes to:
\begin{align}
    \ln{G}_{n_0}(z,t)=\ln{G}(z,t)+n_0\ln{((z-1)e^{-t}+1)}.
\end{align}
Observe that the $k$-th derivative of $\ln{G(z,t)}$ in $z=0$ is given by:
\begin{align}
    \frac{\partial^{k}}{\partial z^{k}}\ln{G}\Big |_{z=0}&=k!(\lambda\ast Q_k)(t),
\end{align}
where
\begin{align}
    Q_k(t)=\sum_{m=0}^\infty q_m\mathcal{B}_k^m(t)
\end{align}
is the probability that a single burst results in $k$ molecules remaining after time $t$.
From this, the probability distribution follows as:
\begin{align}
    p_0(t)&=\exp{\Big((\lambda\ast Q_0)(t) -\int_0^t\text{d}s\lambda(s) \Big )}\ ,\\ \label{eq:sol-p0}
    p_n(t)&=\frac{1}{n!}\frac{\partial^n}{\partial z^n}G(z,t)\Big |_{z=0}\\&=\frac{1}{n!}B_n(1!(\lambda\ast Q_1)(t),...,n!(\lambda\ast Q_n)(t))p_0(t),
\end{align}
where $B_n(x_1,...,x_n)$ denotes the Bell polynomials \cite{comtet2012advanced}, which satisfy the following recurrence relation:
\begin{align}
    B_{n+1}(x_1,...,x_{n+1})=\sum_{i=0}^n\binom{n}{i}B_{n-i}(x_1,...,x_{n-i})x_{i+1}.
\end{align}
This implies the following recurrence relation for $p_n(t)$:
\begin{align}
    p_{n}(t)=\frac{1}{n} \sum_{k=1}^{n}kp_{n-k}(t)(\lambda\ast Q_{k})(t), \label{eq:sol-pn}
\end{align}
which allows for an efficient evaluation of the distribution. Fig.~\ref{fig:histograms} shows an example plot for $\lambda=\text{const}$.

Eq.~(\ref{eq:sol-pn}) makes explicit that the dynamics are governed by two ingredients: the statistics of burst arrivals and the subsequent decay, encoded in the kernels $Q_k(t)$, which represent the effective contribution of bursts generating $k$ molecules that survive up to time $t$, leading to a compact recursive relation between the occupation probabilities at all times.

\new{If $\lambda$ is constant, corresponding to a homogeneous Poisson process and denoted as (P), one may take the limit $t \to \infty$ of Eq.~(\ref{eq:sol-pn}) and obtain the following steady state probabilities (Supplement IB):
\begin{equation}
\begin{split}
   \pi_0^{(P)} &=\exp(-\lambda \sum_{m=1}^{\infty}q_mH_m)\ , \\
   \pi_n^{(P)} &= \frac{\lambda}{n} \sum_{k=0}^{n-1} \pi_k \sum_{l=n-k}^{\infty} q_l\ ,
\end{split}
\end{equation}
where $H_m:=\sum_{k=1}^m\frac{1}{k}$ are the harmonic numbers. From the cumulant generating function $K(x)=\ln G(e^x,t\!\to\!\infty)$, the steady-state cumulants follow as:
\begin{equation}
    \kappa_{\pi,l}^{(P)} =\lambda \sum_{m=1}^{\infty} q_m \sum_{k=1}^{m} k^{l-1}\ . \label{eq:P-steady-state-cumulants}
\end{equation}}

\section{First passage to a threshold}\label{sec:FPTs}

\subsection{Conditional Hitting Probability}

\new{As illustrated by the results in the previous section, the stochastic dynamics separate intrinsically into the statistics of release events and molecular degradation in between. We now use this to characterize the activation of downstream processes as a result of the molecule count exceeding a given threshold. The first-passage time probability density, i.e. the distribution of times when a certain threshold is crossed for the first time, is only accessible via stochastic sampling methods (examples below). As a simpler yet insightful quantity, we introduce the probability $p^+_{n|n_0}(t)$ that the next release will result in a molecule count of $n$, given that the previous release at time $t$ brought the  count to $n_0$. We refer to this quantity as the \textit{conditional hitting probability} or \textit{conditional post-release distribution}. Intuitively, it allows us to analyze the hitting of thresholds not on a continuous-time basis, but on a per-release basis: rather than tracking exactly when the molecule count crosses a threshold, we ask for the probability of a given release event to bring the count above it. We can rewrite $p^+_{n|n_0}(t)$ as the discrete convolution
\begin{equation}
    p^+_{n|n_0}(t) = \sum_{j=0}^{n}q_{n-j} p^-_{j|n_0}(t) \label{eq:conditional-hitting-prob}, 
\end{equation}
of the release size distribution $q$ introduced in Section~\ref{subsec:model} and the \textit{conditional pre-release distribution} $p^-_{|n_0}(t)$, i.e. the probability of having $n$ molecules right before the release. Between release events, molecules degrade independently with rate unity, meaning that the molecule count follows a linear decay process, starting at time $t$ with count $n_0$. Given that a time period $\tau$ passes between releases, the number of remaining molecules then follows a binomial distribution with $N=n_0$ and $p=e^{-\tau}$, again denoted as $\mathcal{B}^{n_0}_j(\tau)$. Taking into account that $\tau$ is randomly distributed according to a distribution $f(\tau)$, we find
\begin{equation}
\begin{split}
    p^-_{j|n_0}(t) &=\int_{0}^{\infty}d\tau\; f(t+\tau)\; \mathcal{B}^{n_0}_j (\tau)  \\
    &:=\left\langle \mathcal{B}^{n_0}_j (\tau)\right\rangle_{\tau\sim f(t)}. \label{eq:pre-release-prob-general}
\end{split}
\end{equation}
If, as assumed previously in Sections~\ref{subsec:model} and \ref{sec:distribution}, the release events follow a time-inhomogeneous Poisson process with rate $\lambda(t)$, we have 
$f(t) =\lambda(t) \exp \left( -\int_{0}^{t} \lambda(s)ds\right)$.
From the convolution in Eq.~(\ref{eq:conditional-hitting-prob}), we can read that the mean molecule count after the next release is
\begin{equation}
    \mu_{+|n_0} (t)= \langle m \rangle +  n_0 \langle e^{-\tau} \rangle_{\tau \sim f(t)},  \label{eq:post-release-mean-general} 
\end{equation}
where $\langle m \rangle$ is the average size of the incoming burst and $n_0 \langle e^{-\tau} \rangle_{\tau \sim f(t)}$ is the expected number of molecules that survived up to this point. Similarly, also all higher cumulants of the post-release distribution $p^+_{|n_0}(t)$ decompose into two separate contributions from the burst size and the degradation process. 
A particularly useful case, illustrated in the following example, arises when the conditional hitting probability $p^+_{|n_0}(t)$ loses its time dependence, enabling the analysis of the steady state and the mean recurrence time.}

\subsection{Homogeneous Poisson Train and Fixed-Interval Train}\label{subsec:fixed_and_Poisson_train}

\begin{figure*}[p]
    \centering
    \includegraphics[width=\textwidth]{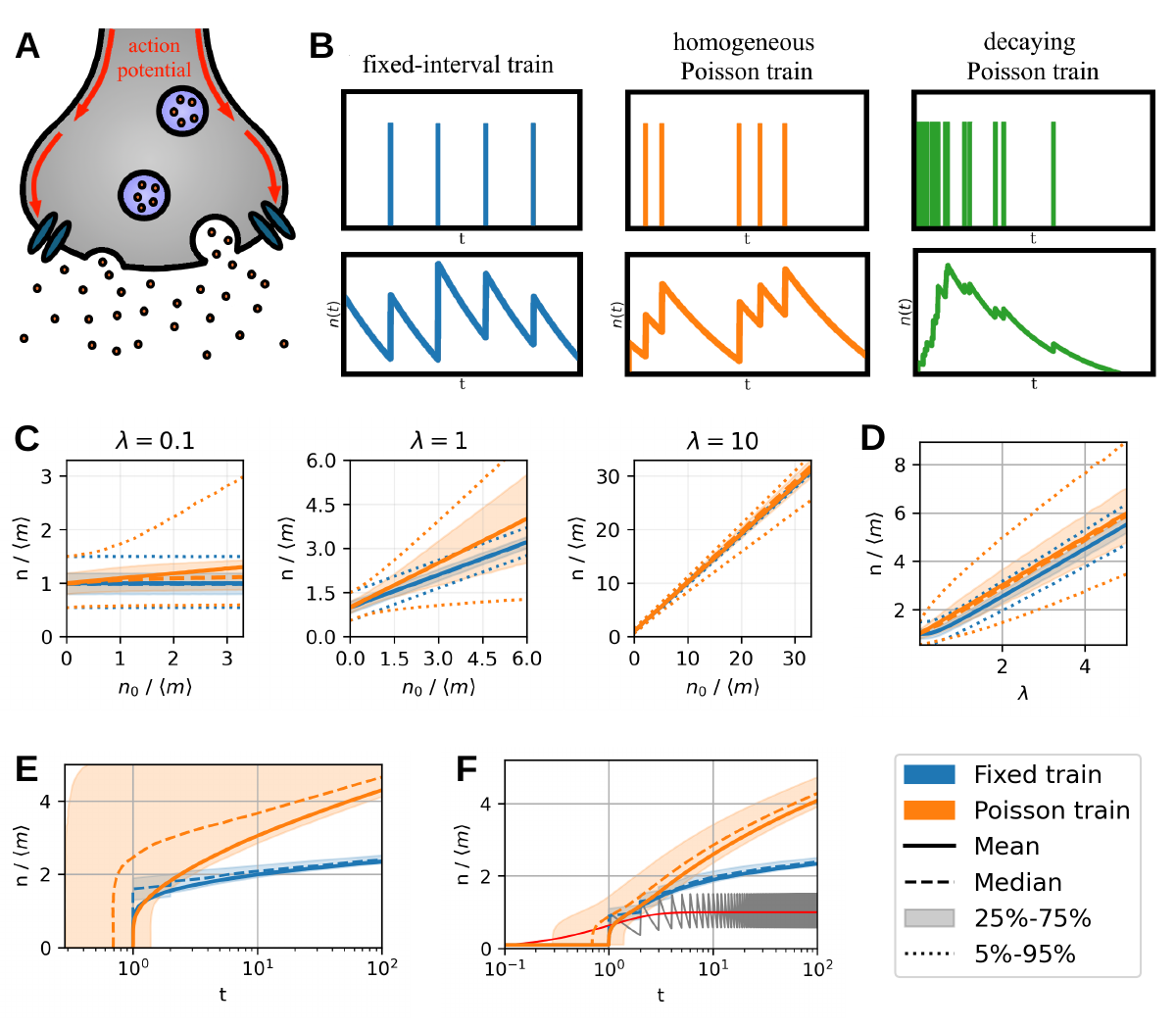}
    \caption{ \new{Comparison of the hitting statistics of a fixed-interval spike train and a homogeneous Poisson spike train with the same mean release rate $\lambda$. The count of released vesicles is assumed to follow a binomial distribution $q^{(V)}=\text{Binom}(V,\rho)$ with $V=50$ releasable vesicles and release probability $\rho=0.2$. The quantal size distribution $q^{(C)}$ is normal with mean $\langle c \rangle=1000$ molecules per vesicle and $\sigma_c=0.2\langle c \rangle$. The quartiles of the shown distributions were obtained from stochastic sampling (see Supplement III). \textbf{A} Upon arrival of a spike, each docked vesicle at the active zone fuses and releases its contents with a certain probability $\rho$. \textbf{B} Qualitative comparison of the trajectories, in the case of releases occurring in fixed intervals (blue) or according to a homogeneous Poisson process (orange). The inhomogeneous Poisson process with exponentially-decaying rate (green) is treated separately in Section~\ref{subsec:decaying_train}.  \textbf{C} Conditional post-release distribution $p^+_{|n_0}$ of the molecule count (normalized by the mean release size $\langle m \rangle$), depending on the previous post-release count $n_0$. \textbf{D} Steady state post-release distribution $\pi^+$  plotted against the release rate $\lambda$. \textbf{E} Distribution of times between subsequent hits of the same threshold (recurrence time), for $\lambda = 1$. Note that a hit is also recorded if the process crosses the threshold and remains above it until the next jump. \textbf{F} First-passage time from $n_0 =0$ to $n$, for $\lambda=1$. For comparison, the transient of the mean molecule count is plotted in gray for the fixed-interval train, and in red for the Poisson train.} }
    \label{fig:steady-state}
\end{figure*}

As a simple and illustrative example, we discuss the release dynamics of a spike train of action potentials at a constant mean rate $\lambda$. \new{Fig.~\ref{fig:steady-state}A schematically
shows the situation as an action potential arrives at a synapse.} Two limiting cases are particularly relevant \cite{dayan2005theoretical, borgers2017introduction}: \new{First, the time-homogeneous Poisson train (P) introduced in Section~\ref{sec:distribution}, where releases are uncorrelated in time, which may apply to spontaneous or irregular neuronal activity. Second, the fixed-interval train (F), in which stimulation is perfectly periodic with inter-spike interval $\frac{1}{\lambda}$, as is typical in controlled experimental settings. Example trajectories of both cases are shown in Fig.~\ref{fig:steady-state}B. Note that the master equation treatment in Sections~\ref{subsec:model} and \ref{sec:distribution} implicitly assumes events to follow a (time-inhomogeneous) Poisson process and therefore does not apply to the fixed-interval train. Instead, in Supplement ID and IE, we derive formulas for the steady-state dynamics using an alternative generating function technique. 

Here, we again focus on the first-passage problem. We begin by revisiting the conditional pre-release distribution $p^-_{|n_0}$ as introduced in Eq.~(\ref{eq:pre-release-prob-general}). For the fixed train, we simply have a binomial distribution with $N=n_0$ and $p=e^{-1/\lambda}$:
\begin{equation}
    p^{-(F)}_{j|n_0} = \mathcal{B}^{n_0}_j (1/\lambda). \label{eq:p--F}
\end{equation}
For the Poisson train, we instead find 
\begin{equation}
\begin{split}
    p^{-(P)}_{j|n_0} &= \int_{0}^{\infty} \lambda e^{-\lambda \tau} \;\mathcal{B}^{n_0}_j(\tau) \; d\tau\\
    &=\lambda \binom{n_0}{j} \,\text{Beta}(j+\lambda, n_0-j+1), \label{eq:p--P}
\end{split}
\end{equation}
corresponding to a $\text{Beta-Binomial}(n_0,\lambda,1)$ distribution. The cumulants of both distributions are well known, and according to Eq.~(\ref{eq:conditional-hitting-prob}), adding to them the cumulants of the release size distribution $q$ yields the cumulants of the post-release distribution $p^+_{|n_0}$. For example, we find for mean $\mu_{+|n_0}$ and variance $\sigma_{+|n_0}^{2}$:
\begin{align}
\mu_{+|n_0}^{(F)} &= n_0e^{-1/\lambda} + \langle m \rangle,  \\
{\sigma_{+|n_0}^{2(F)}} &= n_0e^{-1/\lambda}(1-e^{-1/\lambda}) + \sigma_m^2, \\
\mu_{+|n_0}^{(P)} &=  n_0 \frac{\lambda}{\lambda +1} + \langle m \rangle, \\
    {\sigma_{+|n_0}^{2(P)}} &= n_0\frac{\left(\frac{n_0}{\lambda +1} +1\right)\lambda}{(\lambda+1)(\lambda+2)}+  \sigma_m^2, \label{eq:sigma-p-P-n0}
\end{align}
where $\langle m \rangle$ is the mean and $\sigma_m^2$ is the variance of the release size. In particular, the mean post-release count $\mu_{+|n_0}$ is strictly smaller for the fixed train than for the Poisson train, despite the release rate $\lambda$ and the release size being the same. Note that this is a direct consequence of the different decay probabilities between releases, as described by Eqs. (\ref{eq:p--F})  and (\ref{eq:p--P}). Similarly, the variance $\sigma_{+|n_0}^{2}$ scales as $O(n_0)$ for the fixed train and as $O({n_0}^2)$ for the Poisson train, implying substantially greater post-release variability from the Poisson train when the count is far above the steady-state mean. In Fig.~\ref{fig:steady-state}C we plot the conditional post-release distributions for different rates $\lambda$, including the median and quartiles to illustrate the different shapes of the distributions. $p_{|n_0}^{+(P)}$ appears skewed for low and high release rates, while $p_{|n_0}^{+(F)}$ is approximately normal (see Supplement IE for a formal argument on why this is the case).

By interpreting $p_{n|n_0}^{+}$ as a transition probability of a discrete-time stochastic process (Supplements IC and ID), we obtain analytical expressions for the cumulants of the post-release distribution in steady state $\pi^+$. In other words, we examine the distribution $\pi^+$ of the molecule count after a release without making an assumption on the molecule count $n_0$ after the previous release. In Fig.~\ref{fig:steady-state}D, we plot the mean and quartiles of $\pi^+$ over the release rate $\lambda$. We observe that the findings for the conditional distribution essentially also apply to the steady state, with the post-release mean for the fixed train
\begin{equation}
    \mu^{(F)}_+ = \frac{\langle m \rangle}{1-e^{-1/\lambda}} \label{eq:mu-F-p}
\end{equation}
being strictly lower than the post-release mean of the Poisson train
\begin{equation}
    \mu^{(P)}_+ = (\lambda +1)     \langle m \rangle \label{eq:mu-P-p}
\end{equation}
and the Poisson train showing a higher variance. 

With the steady state results for the post-release distribution, we turn to the recurrence time to a threshold, defined as the waiting time between successive release events at which the molecule count reaches or exceeds a threshold $n$. According to a standard theorem in  Markov chain theory \cite{norris1998markov}, the mean number of release events needed to reach a molecule count equal to or higher than $n$ is $1/\pi^+_{\ge n}$, where $\pi^+_{\ge n} = \sum_{k=n}^{\infty} \pi^+_k$. Since the mean time between release events is $1/\lambda$, the mean recurrence time is $\langle \tau_{n} \rangle = 1/(\lambda \pi^+_{\ge n})$. As shown in Fig.~\ref{fig:steady-state}E, this time grows super-exponentially with threshold $n$ and is generally higher for the fixed-interval train compared to the Poisson train, reflecting the lower mean and variance in the post-release count. As for the post-release distribution, the variability in inter-event timing causes the recurrence time distribution to be broader for the Poisson train than for the fixed-interval train. 

The same super-exponential growth with threshold, and the contrast between Poisson and fixed-interval trains, holds qualitatively for the first-passage time from any lower initial count. As an example, consider a neuron that is initially inactive $(n_0=0)$ and begins receiving stimulation at $t=0$. The first-passage time is then the time for the molecule count to first reach a threshold  value $n$. As a reference, we compare the first-passage time distribution against the transient evolution of the mean molecule count.
For the Poisson train, we apply Eq.~(\ref{eq:me-mean-evolution}) and find
\begin{equation}
    \langle n^{(P)}\rangle(t) = \langle m\rangle \lambda (1-e^{-t}).
\end{equation}
For the fixed-interval train, the mean molecule count after the $N$-th release event is $\langle n \rangle_N = \langle m\rangle +e^{-1/\lambda} \langle n \rangle_{N-1}$, where $\langle n \rangle_0 =0$ and each step accounts for degradation between releases. Solving the recursion and including exponential decay between releases yields
\begin{equation}
    \langle n^{(F)}\rangle(t) =  \langle m \rangle  \frac{1- e^{-\lfloor \lambda t \rfloor/\lambda}}{1-e^{-1/\lambda}} e^{-T},
\end{equation}
where $T=t- \lfloor \lambda t \rfloor/\lambda$ is the time since the last release. In Fig.~\ref{fig:steady-state}F, we compare this to the sampled first-passage time distribution. The mean first-passage time curve (blue and orange) and the transient mean molecule count (red and grey) intersect at a threshold value below the steady-state mean. For low thresholds, the first-passage time exceeds the transient mean time, since some finite time must elapse before the first jump occurs in the stochastic setting. For high thresholds, the stochastic process reaches the threshold earlier and the mean first-passage time approaches the mean recurrence time (Fig.~\ref{fig:steady-state}E), as the influence of the initial condition diminishes. We also observe that both the recurrence and the first-passage time distributions for high thresholds converge to a memoryless exponential distribution. This likely reflects that many individual attempts have to be made to cross a threshold, and has been rigorously proven in the case of exponential release size distributions \cite{bremont2026beyond}.

The results above for the Poisson and fixed-interval trains illustrate how stochasticity in both release size and release timing shapes the activation of downstream processes. Beyond that, they provide the foundation for more elaborate models. In Supplement IC and IIC, we generalize the results for the pre- and post-release distribution to the case of gamma-distributed inter-arrival times, which may be used to model bunched or anti-bunched spike trains. Section~\ref{sec:vesicle-depletion} extends the results to changing vesicle-pool sizes. }

\subsection{Poisson Train with Exponentially Decaying Release Rate}\label{subsec:decaying_train}

\begin{figure*}[t]
    \centering
    \includegraphics[width=\textwidth]{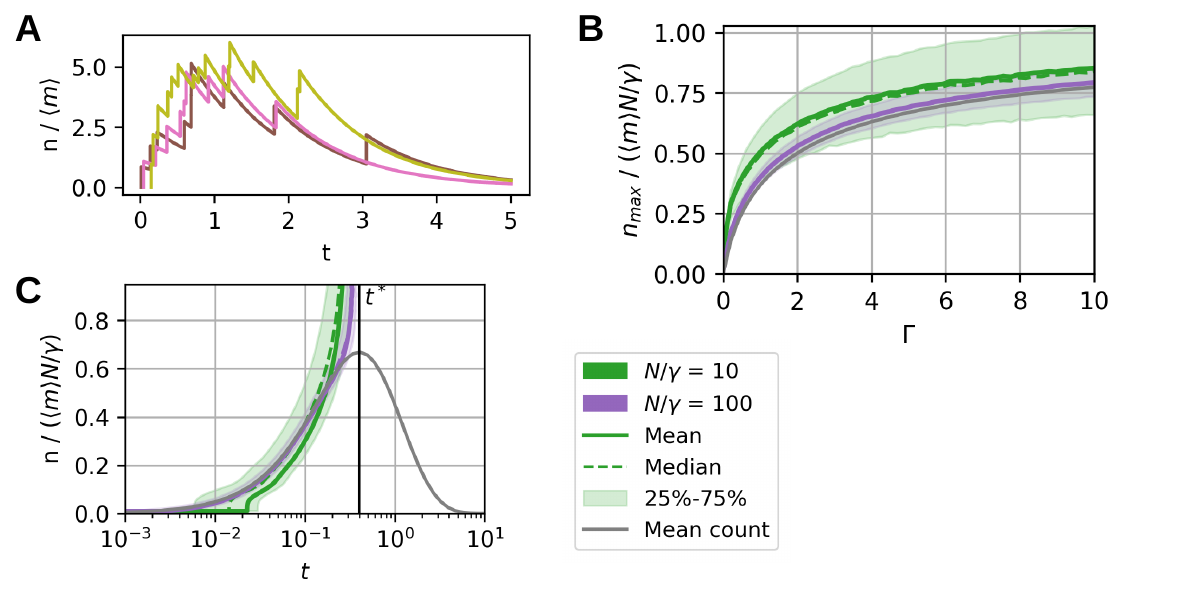}
    \caption{\new{Hitting statistics of a time-inhomogeneous Poisson spike train, where the release rate decays exponentially with rate $\Gamma$ and the mean total number of release events is $N$. The release size distribution is the same as in Fig.~\ref{fig:steady-state}. \textbf{A} Example trajectories for $N/\gamma =10$ and $\Gamma =1$, with the molecule count $n$ normalized by the mean release size $\langle m \rangle $. \textbf{B} Distribution of maximum threshold reached, compared to the "deterministic" maximum of the mean molecule count in gray. \textbf{C} First-passage time from a count of $0$ to threshold $n$, provided that the threshold is reached, compared to the transient of the mean molecule count in gray ($\Gamma =5$). The time $t^*$ when the mean molecule count reaches its maximum is significantly later than the mean first-passage time. }}
    \label{fig:decaying-neuron}
\end{figure*}

As a second example, we study a case in which the release rate is time-dependent. Neuronal activity often occurs in transient bursts, for example, following a stimulus or during short periods of elevated excitability \cite{dayan2005theoretical, borgers2017introduction}. \new{To capture this behavior, we consider a neuron that is inactive for $t <0$ and for $t>0$ releases vesicles according to a time-inhomogeneous Poisson process with rate $\kappa(t)= N\,\Gamma e^{-\Gamma t}$, where $N = \int_0^{\infty} \kappa(t) dt$ is the mean total number of release events and $\Gamma$ is a relaxation rate. To avoid vesicle depletion (see  Section~\ref{sec:vesicle-depletion}), we assume $\kappa(t)$ to be much smaller than the rate of vesicle replenishment.  As in Section~\ref{subsec:model}, we can non-dimensionalize the release rate by rescaling time as $\gamma t \rightarrow t$, where $\gamma$ is the rate of degradation of molecules, and introduce $\lambda(t) = \frac{\kappa(t)}{\gamma} $.} 

Fig.~\ref{fig:decaying-neuron}A shows example trajectories of the molecule count for this case\new{, and Fig.~\ref{fig:steady-state}B compares them qualitatively to the fixed-interval and the time-homogeneous Poisson train. From Eq.~(\ref{eq:me-mean-evolution}), we find 
\begin{equation}
    \langle n(t) \rangle = \frac{\langle m \rangle N }{\gamma} \frac{\Gamma(e^{-t}-e^{- \Gamma t})}{\Gamma -1}
\end{equation}
for the time evolution of the mean. At $t^* = \ln(\Gamma)/(\Gamma-1)$, this becomes maximal with $\langle n(t^*) \rangle = \langle m \rangle \frac{N}{\gamma} \exp(-\ln\Gamma /(\Gamma-1))$. In Fig.~\ref{fig:decaying-neuron}B, we compare $\langle n(t^*) \rangle$ to the mean of the actual maximum molecule count reached by the stochastic process. For all relaxation rates $\Gamma$, the true stochastic maximum is strictly higher, particularly when $N / \gamma$ is small, which corresponds to the case of few release events or fast degradation.} In Fig.~\ref{fig:decaying-neuron}C, we compare the transient of the mean molecule count to the distribution of the first-passage time to a threshold, given that the threshold is actually reached. We observe that thresholds around $\langle n(t^*) \rangle$ and higher thresholds are in fact reached significantly before $t^*$. For lower thresholds, the dynamics of the mean are only representative \new{if $N / \gamma$ is large. Together, these results highlight that the mean molecule count can be  misleading in the case of few release events or fast degradation, underestimating the maximum count reached and overestimating the time until a threshold is crossed.}

\section{Vesicle Depletion}\label{sec:vesicle-depletion}

\begin{figure}[t]
    \centering
    \includegraphics[width=0.6\linewidth]{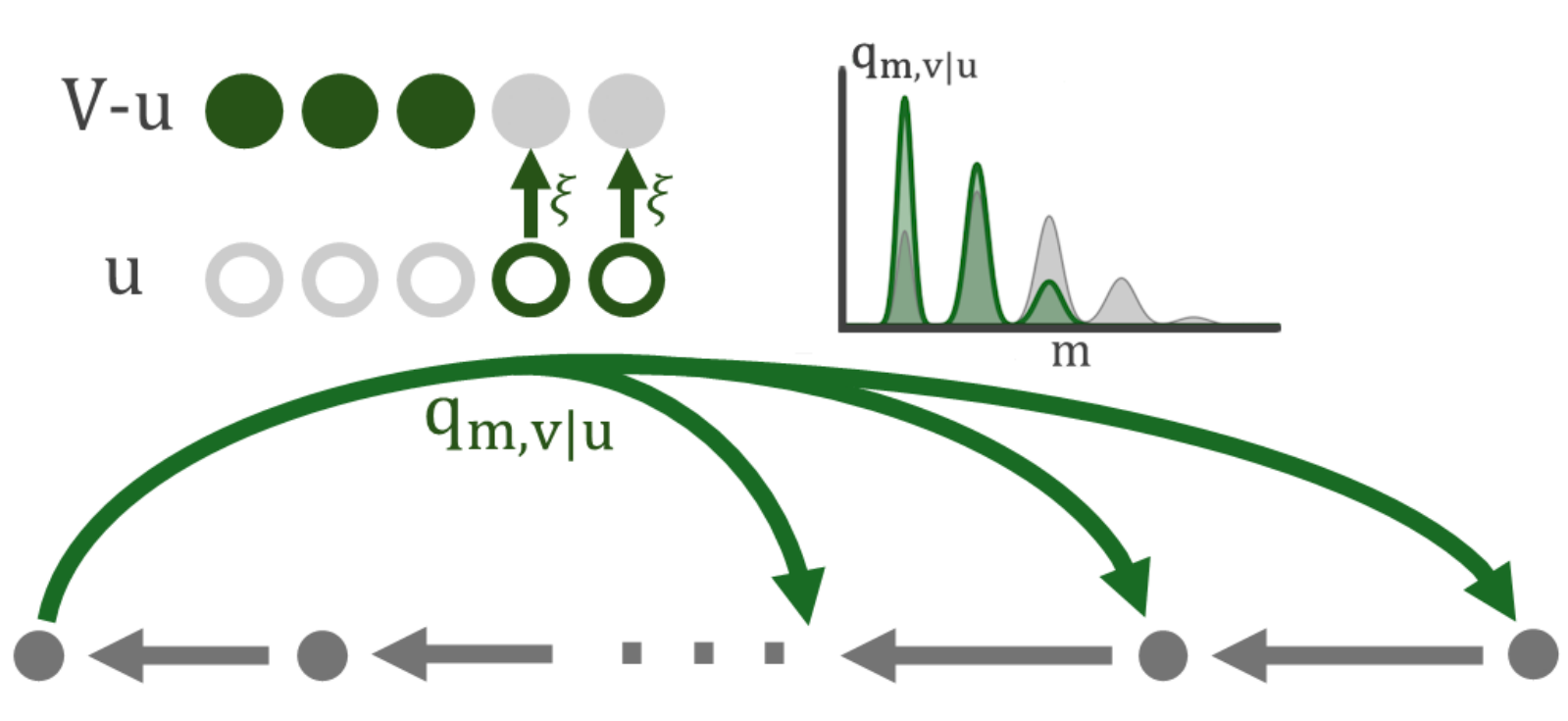}
    \caption{Extended model including the effects of vesicle depletion. Vesicles are released from a finite pool of $V-u$ occupied sites that replenish over time at a rate $\xi$. The burst size distribution is no longer constant in this case but depends on $u$.}
    \label{fig:depletion_model}
\end{figure}

For high release rates, the pool of releasable vesicles can be depleted, and the mathematical description therefore requires a release size distribution $q$ depending on previous release events. Let $V$ be the total number of release sites and $u$ the number of empty release sites. Following previous work by other authors \cite{pulido2015vesicular, rosenbaum2012short, loebel2009multiquantal, fuhrmann2002coding, rijal2024exact,ali2025exact, gambrell2024feedforward, gambrell2025analysis}, we assume a constant replenishment rate $\xi$ per released vesicle (see Fig.~\ref{fig:depletion_model})\new{, which we again non-dimensionalize by applying $\xi /\gamma \to \xi $, where $\gamma$ is the molecular degradation rate}.  

The extended version of Eq.~(\ref{eq:master_equation}) then reads
\begin{equation}
\begin{split}
    &\dt{p_{n,u}}= -( n+\lambda+ \xi u)p_{n,u}+ (n+1)p_{n+1,u} + \xi (u+1)p_{n,u+1} + \lambda\sum_{m,v} q_{m,v|u-v} p_{n-m,u-v},\label{eq:master_equation_depletion}
\end{split}
\end{equation}
where $q_{m,v|u_0}$ now is the probability of releasing $m$ molecules distributed across $v$ vesicles, provided that $V-u_0$ vesicles are available for release. In accordance with the binomial model, we choose
\begin{equation}
    q_{m,v|u_0}=\left(q^{(C)}\right)^{*v}_m \text{Binom}(v \;|\;V-u_0, \rho),
\end{equation}
where $q^{(C)}$ is the quantal size distribution with mean $\langle c \rangle$ and variance $\sigma_c^2$. We note that a simpler version of this model, where the quantal size is fixed to $\langle c \rangle$, has been covered in recent work by Rijal et al. \cite{rijal2024exact} and other authors \cite{gambrell2024feedforward, gambrell2025analysis}. In Supplement II, we extend their moment and generating function analysis to account for a non-vanishing release size variance $\sigma_c^2$ and provide generalized formulas for the mean molecule count and the post-release distribution.

\begin{figure*}[t]
    \centering
    \includegraphics[width=\textwidth]{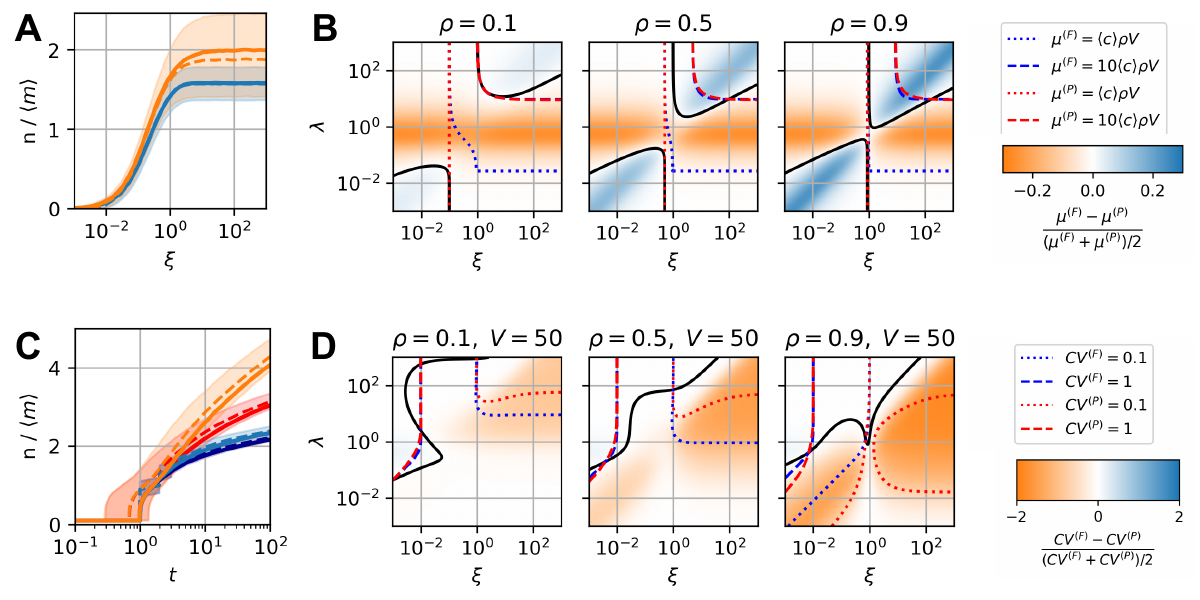}
    \caption{Fixed-train (blue) and the Poisson train (orange), if vesicles are replenished with some finite rate $\xi$. \textbf{A} Post-release steady state distribution at release rate $\lambda =1$ and release probability $\rho = 0.2$, as a function of $\xi$. Like in  Fig.~\ref{fig:steady-state}, we plot mean, median and the 25\% and 75\% quartiles, which were obtained from stochastic sampling with \new{mean quantal size $\langle c \rangle=1000$, $V=50$ release sites and quantal size variation $\sigma_c=0.2\langle c \rangle$}. \textbf{B} Mean post-release counts for different \new{release probabilities $\rho$, mean release rates $\lambda$ and replenishment rates $\xi$}. The black lines indicate the choice of parameters for which the means for the fixed train and for the Poisson train coincide. The dashed contour lines mark specific values (see legend).  \textbf{C} First-passage time distributions for $\lambda =1$, $\rho=0.2$, $\langle c \rangle=1000$, $V=50$ and $\sigma_c=0.2\langle c \rangle$. In blue and orange, we plot the distributions for a high rate of replenishment $\xi = 10^3$, in dark blue and red for a low rate $\xi = 1.2$. \textbf{D} Coefficient of variation $CV=\sigma /\mu$ of the post-release distributions, plotted following the same scheme as in B. }
    \label{fig:depletion}
\end{figure*}

The steady-state distributions for both the fixed-train and the Poisson train (cf. Section~\ref{subsec:fixed_and_Poisson_train}) can change substantially under the influence of vesicle depletion. For example, Fig.~\ref{fig:depletion}A shows that for $\lambda=1$ and even a low release rate of $\rho =0.2$, the mean and variance of the post-release distribution  are significantly altered if the replenishment rate $\xi$ is on the order of magnitude of the release rate $\lambda$. This is also true for the first-passage times to thresholds above the steady state mean, which are plotted in Fig.~\ref{fig:depletion}C.

In Supplement IIC, we derive the mean and variance of the modified post-release distributions \new{$p^+$}. It turns out that the previous Eqs.~(\ref{eq:mu-F-p}) and (\ref{eq:mu-P-p}) for the steady-state mean \new{$\mu_+$} with constant vesicle pool remain valid, but only if the mean release size $\langle m \rangle$ is treated as a function of the release rate $\lambda$ and the rate of replenishment $\xi$. For the fixed train, one finds
\begin{equation}
    \langle m^{(F)} (\lambda,\xi)\rangle = \langle c \rangle V \rho  \cdot \frac{1 - e^{-\xi / \lambda}}{1-(1-\rho)e^{-\xi /\lambda}},
\end{equation}
while for the Poisson train
\begin{equation}
    \langle m^{(P)} (\lambda,\xi)\rangle = \langle c \rangle V \rho \cdot \frac{\xi}{\xi + \lambda \rho},
\end{equation}
which is always smaller than the value for the fixed train.

As a consequence, the steady state mean may be higher either for the fixed train or for the Poisson train, depending on the choice of $\lambda$, $\xi$ and $\rho$. In Fig.~\ref{fig:depletion}B,  we plot the difference between the respective means \new{$\mu_+$} of the post-release distribution for different parameter choices. As indicated by the dashed contour lines, the mean generally increases with both $\lambda$ and $\xi$. In a region where $10^{-1} \lesssim \lambda \lesssim 10^1$ (orange shade), the Poisson train tends to give a higher mean, while in a region $\lambda \approx \xi$ (blue shade), it is the fixed train. Where both regions intersect (black line), it depends on the release probability per vesicle $\rho$, with high $\rho$ favoring the fixed train. \new{This makes intuitive sense, since the Poisson train yields higher means for a given release size (cf. Section~\ref{subsec:fixed_and_Poisson_train}), especially for intermediate $\lambda$, but also exhibits clustering of release events and therefore lower mean release sizes compared to the fixed train.} This matters especially for high $\rho$, i.e., if many of the available vesicles are released.

If the release sizes are generally small ($\lambda > \xi$ and small $V$), the variability of the post-release distribution, quantified by the coefficient of variation $\sigma /\mu$, may also become near-identical between fixed train and Poisson train or even slightly higher for the fixed train. Otherwise, as seen in Fig.~\ref{fig:depletion}D for the example $V=50$, the variability of the Poisson train is significantly higher. 

\section{Summary and Discussion}
\label{summary}

In this work, we have derived an exact solution for the fully time-dependent occupation probabilities of a general batch arrival-degradation model and shown that the activation timing is not determined by the mean burst frequency alone, but also by the time statistics of the bursts. Our model for vesicle-mediated signaling combines the effects of burst-like releases at generally time-dependent rates, arbitrary release size distributions and molecular degradation. By solving the corresponding master equation with the generating function approach, we obtained an exact representation of the full time-dependent probability distribution that could be used to derive an interpretable recurrence relation to evaluate the dynamics. 

To study the activation of downstream processes, such as the activation of a postsynaptic neuron or a hormone-sensing target cell, we introduced the conditional threshold crossing probability and studied the first-passage time to a prescribed target for multimodal burst size distributions reflecting the effect of multi-vesicle releases. We compared the two experimentally relevant cases of a fixed spike train and a homogeneous Poisson spike train with the same mean vesicle-release frequency, reflecting the cases of periodic and irregular excitation of the presynaptic neuron, respectively. Our results show that the two activation profiles show qualitatively distinct first-passage time behaviors, due to different fluctuations in the timing of release events determining the probability of rare fluctuations that result in sufficiently high molecule counts for threshold crossing. In particular, this illustrates that the mean vesicle-release frequency is insufficient to fully capture activation profiles. 
Moreover, this effect also persists for time-dependent release rates, such as an exponentially decaying activity after an initial stimulus, where threshold crossing is dominated by early stochastic events and can occur significantly before the maximum of the mean molecule count.

Lastly, we also investigated an extension of the model that incorporates vesicle depletion due to a finite replenishment rate. In this extended model, the number of available vesicles becomes a dynamical variable coupling the burst size distribution to previous release events. Qualitatively, the influence of stochasticity persists, but depletion can alter the differences between activation profiles, depending on the replenishment rate and the release probability. This demonstrates that burst statistics have a robust influence that nevertheless depends on the conditions of the system.

\new{Including the number of available vesicles as a second stochastic variable introduces memory to the system. Without it, the burst-events are Poisson-like, i.e., memoryless, which is only valid for high replenishment rates of the vesicles ready for release. This assumption can also be relaxed by allowing for non-exponentially distributed release times, which we discuss for the analytically tractable case of a gamma distribution of the release times in Supplement IC and IIC for the simple model and the model including vesicle depletion, respectively.}

Our results complement and extend previous work on stochastic burst-degradation processes and first-passage times. While earlier studies have largely focused on steady state distributions or specific burst size distributions such as geometric distribution in gene expression \cite{friedman2006linking, shahrezaei2008analytical, rijal2022exact, wang2025joint}, our framework allows for arbitrary release statistics and time-dependent burst rates, which are both reflected in our analytical expression for the time-dependent occupation probabilities. Recent work has shown that the prediction of mean activation times for noisy systems can differ profoundly from deterministic expectations \cite{ham2024stochastic}, \new{a fact that is also well-known for diffusion problems in complex environments, where the means fail to capture the broadness or heterogeneity of the FPT distribution~\cite{mattos2012first,benichou2014first,godec2016first}}. Extending this idea, our first-passage time analysis also reveals that different stochastic systems such as the fixed-interval and homogeneous Poisson train vesicle release models can lead to distinct activation behavior despite sharing the same mean burst frequency. Similar qualitative results were obtained for time-dependent release rates and also under the inclusion of vesicle depletion, indicating that the statistical nature of release events constitutes a key parameter of activation dynamics.

From a biological perspective, these findings suggest that variability in neuronal or endocrine signaling may not merely reflect a noisy environment but could actively shape downstream processes, and irregular or burst-like activation patterns could enhance responsiveness by increasing the probability of fast activation, while vesicle depletion provides a mechanism to regulate this effect.

So far, our model assumes independent release events and linear degradation, neglecting feedback mechanisms, spatial transport and receptor dynamics  \cite{franks2003independent, gambrell2025analysis, gambrell2024feedforward}. Extending the framework to include such effects and potentially interacting signaling pathways may provide further insight into how stochasticity shapes cellular communication in realistic biological settings. More broadly, our framework suggests \new{that asymmetric dynamics on} the microscopic level give rise to emergent activation dynamics at the macroscopic scale. 

\new{Finally, our analysis was restricted to a single type of molecule with a single activation threshold, while in practice vesicle releases often feature more than one species, which may serve as different downstream signals \cite{nusbaum2017functional}. In the future, extensions of our mathematical model to $d$-dimensional vectors of molecular counts for $d$ different species may allow to further explore correlations between their individual first-passage times. For molecular species that are co-packaged, these correlations are expected to be particularly strong and future research may explore how correlations change due to individual degradation and diffusion patterns, including in comparison to conventional diffusion.}

%
% Each of the commands below will create an unnumbered section with the appropriate heading.
% Remove any sections that are not relevant for your article.
% All sections except suppdata will be removed if the [anonymous] option is used.
% See iopjournal-guidelines.pdf for more information.
%

\ack{JBV thanks the German Academic Scholarship Foundation (Studienstiftung des Deutschen Volkes) for support. We also acknowledge support by the Max Planck School Matter to Life funded by the Dieter Schwarz Foundation and the Max Planck Society.}

%\roles{Sample text inserted for demonstration.}
% List author names and the contributions made to the article, using terms from the NISO Contributor Roles Taxonomy (CRediT) https://credit.niso.org

\data{All data that support the findings of this study are included within the article and the supplementary files. The code used to generate the figures is available at \url{https://github.com/HaukeJan/vesicle-release}.}

% For more information on IOP Publishing's research data policy see: https://publishingsupport.iopscience.iop.org/questions/research-data/

%\suppdata{Sample text inserted for demonstration.}

%\section*{References}

%\bibliography{apssamp}

\providecommand{\newblock}{}

\end{document}

\title{Supplemental Material for\\
Activation in Vesicle-Mediated Signaling Shaped by Batch Arrival Statistics}

\author{Jan Hauke\textsuperscript{1}}
\author{Julian B. Voits\textsuperscript{1}}
\author{Ulrich S. Schwarz\textsuperscript{1,2}}%
 \email{Corresponding author: schwarz@thphys.uni-heidelberg.de}
\affiliation{% 
\textsuperscript{1}Institute for Theoretical Physics, Heidelberg University, Germany\\ \textsuperscript{2}BioQuant-Center for Quantitative Biology, Heidelberg University, Germany
}%
%\date{\today}

\maketitle

\tableofcontents
\bigskip

\break

In Section~\ref{sec:appendix-main-model}, we derive results for the main model discussed in the paper, where the molecular count is the only stochastic variable and the release size distribution is constant. In Section~\ref{sec:appendix-extended-model}, we cover the extended version, as introduced in Section 5 of the main text, that accounts for the depletion of vesicles. In Section~\ref{sec:appendix-sampling}, we provide details on how to efficiently obtain samples from the stochastic process.

Throughout the calculations, we distinguish between different cases for the next-arrival distribution of incoming action-potential spike trains. In Table~\ref{tab:terminology}, we give a quick overview over the terminology. 

\begin{table}[h]
\centering
\caption{Terminology of Spike Trains }
\setlength{\tabcolsep}{12pt} % horizontal padding
\renewcommand{\arraystretch}{1.5} % vertical padding
\begin{tabular}{|c|c|c|}
\hline
 & inter-arrival time $\tau$ (at time $t$) & parameters \\ 
\hline
General Train &  $\tau \sim f(t,\tau)$ & pdf $f(t, \tau)$\\ 
General Poisson Train & $\tau \sim \text{Exp}(\lambda(t))$ & rate function $\lambda(t)$  \\ 
Gamma-interval Train & $\tau \sim \text{Gamma}(\alpha, \beta\!=\!\alpha\lambda)$ & rate $\lambda$, shape $\alpha >0$ \\
(Homogeneous) Poisson Train $(P)$ & $\tau \sim \text{Exp}(\lambda)$ & rate $\lambda$ \\ 
Fixed-interval Train $(F)$ & $\tau = 1/\lambda$ & rate $\lambda$ \\ 
\hline
\end{tabular}
\label{tab:terminology}
\end{table}

The primary examples discussed in the main text are the homogeneous Poisson train $(P)$ and the fixed-interval train $(F)$. Note that the homogeneous Poisson train is a special case both of the general Poisson train (if $\lambda = \text{const}$) and of the gamma-interval train (if $\alpha =1$). The fixed-interval train corresponds to the $\alpha \to \infty$ limit of the gamma-interval train.

\section{Main Model \label{sec:appendix-main-model}}

\subsection{\label{sec:appendix-derivation-probability-density} General Poisson Train: Transient Probabilities }

Applying the generating function
\begin{align}
    G(z,t) = \sum_{n=0}^\infty p_n(t) z^n
\end{align}
to the master equation:
\new{\begin{align}
    \partial_tG(z,t)&=-\sum_{n=0}^\infty (\lambda(t) + n)p_n z^n+ \sum_{n=0}^\infty (n+1)p_{n+1} z^n+\lambda(t) \sum_{n=0}^\infty\sum_{m=0}^\infty p_{n-m}q_m z^n,
\end{align}}
results in the following partial differential equation for $G(z,t)$ after rewriting the sums:
\begin{align}
    \partial_tG(z,t)&=-\lambda(t)G(z,t) + (1-z)\partial_zG(z,t) + \lambda(t) \sum_{m=0}^\infty q_m z^m G(z,t)\\ \partial_t\ln{G(z,t)}&=(1-z)\partial_z \ln G(z,t)+ \lambda(t) \Big(\sum_{m=0}^\infty q_m z^m-1\Big)\label{eq:PDE_generating_function}.
\end{align}
There is a steady state solution $G^s(z)$, if the limit $\lambda(t\to\infty)=:\lambda_\infty$ exists. From the condition $\partial_t \ln G^s = 0$, we obtain
\begin{align}
    \partial_z \ln G^s=\lambda_\infty \frac{\sum_{m=0}^\infty q_mz^m - 1}{z - 1}=\lambda_\infty \sum_{m=0}^\infty q_m\frac{z^m - 1}{z - 1}=\lambda_\infty \sum_{m=0}^\infty q_m\sum_{l=0}^{m-1}z^l,
\end{align}
using that $\sum_{m=0}^\infty q_m=1$. Integrating and imposing $G^s(1)=1$ for normalization yields
\begin{align}
    G^s(z)=\exp\Big(\lambda_\infty \sum_{m=0}^{\infty} q_m\sum_{l=1}^m\frac{z^l - 1}{l}\Big)=\exp\Big(\lambda_\infty \sum_{l=1}^\infty\sum_{m=0}^{l} q_m\frac{z^l - 1}{l}\Big),
\end{align}
so the steady state distribution belongs to the class of compound Poisson distributions \cite{wimmer1996multiple}.
The cumulant-generating function is given by
\begin{align}
    K(x)=\ln G^s(e^x)=\lambda_\infty \sum_{l=1}^\infty\sum_{m=0}^{l} q_m\frac{e^{lx} - 1}{l}=\sum_{k=1}^\infty \frac{1}{k!}\big(\lambda_\infty \sum_{l=0}^\infty l^{k}~\sum_{m=0}^{l+1}q_m\big)x^k
\end{align}
Thus, the $k$-th cumulant reads:
\begin{align}
    \kappa_k^s = \lambda_\infty \sum_{l=0}^\infty l^{k}~\sum_{m=0}^{l+1}q_m
\end{align}

For the time-dependent case, Eq.~(\ref{eq:PDE_generating_function}) the solutions are of the following shape \cite{polyanin2025handbook}:
\begin{align}
    \ln G(z,t)=\ln G((z-1)e^{-t},0)+ \int_0^t\text{d}s~\lambda(s)\Big(\sum_{m=0}^\infty q_m\big(1+(z-1)e^{s-t}\big)^m-1\Big)
\end{align}
Imposing the initial condition $p_n(0)=\delta_{n,n_0}$ corresponding to $\ln G(z,0)=n_0\ln z$ yields
\begin{align}
    \ln G(z,t)=n_0\ln((z-1)e^{-t}+1)+ \int_0^t\text{d}s~\lambda(s)\Big(\sum_{m=0}^\infty q_m\big(1+(z-1)e^{s-t}\big)^m-1\Big).
\end{align}
For simplicity, we choose $n_0=0$ in the following calculations. Then, one finds:
\begin{align}
    \ln G(z,t)&=\Big(\sum_{m=0}^\infty q_m\int_0^t\text{d}s~\lambda(s)\big(1+(z-1)e^{s-t}\big)^m\Big)-\int_0^t\text{d}s~\lambda(s)\\ &=\sum_{m=0}^\infty q_m\int_0^t\text{d}s~\lambda(s)\sum_{k=0}^m\binom{m}{k}e^{k(s-t)}(1-e^{s-t})^{m-k}z^k-\int_0^t\text{d}s~\lambda(s)\\ &=\sum_{m=0}^\infty \sum_{k=0}^mq_m\big(\lambda\ast \mathcal{B}^m_k\big)(t)z^k-\int_0^t\text{d}s~\lambda(s)\\&=\sum_{k=0}^\infty\sum_{m=k}^\infty q_m\big(\lambda\ast \mathcal{B}^m_k\big)(t)z^k-\int_0^t\text{d}s~\lambda(s),
\end{align}
identifying the binomial probability for $k$ successes in $m$ trails with probability $e^{-t}$:
\begin{align}
    \mathcal{B}^m_k(t):=\binom{m}{k}e^{-kt}(1-e^{-t})^{m-k}.
\end{align}

The occupation probabilities follow from derivatives of $G(z,t)$ at $z=0$:
\begin{align}
    p_n(t)=\frac{1}{n!} \frac{\partial^n}{\partial z^n} G(z,t)\Big|_{z=0}.
\end{align}
To compute these expressions, we first consider the derivatives of $\ln G(z,t)$:
\begin{align}
    \frac{d^{k}}{dz^{k}}\ln{G}\Big |_{z=0}&=k!(\lambda\ast Q_k)(t),
\end{align}
where
\begin{align}
    Q_k(t)=\sum_{m=0}^\infty q_m\mathcal{B}^m_k(t).
\end{align}
The occupation probability for $n=0$ simply reads:
\begin{align}
    p_0(t)=\exp{\Big((\lambda\ast Q_0)(t) -\int_0^t\text{d}s\lambda(s) \Big )}, \label{eq:p0}
\end{align}
and setting $f(z,t):=\ln{G}(z,t)$ for $n>0$, one can apply Faà di Bruno's formula on $\exp\big(f(z,t)\big)$ to obtain:
\begin{align}
    p_n(t)=\frac{1}{n!}B_n\big(f^{(1)}(z=0,t),\dots,f^{(n)}(z=0,t)\big)\,p_0(t),
\end{align}
where $B_n(x_1,...,x_n)$ denotes the $n$-th exponential Bell-polynomial \cite{comtet2012advanced}. Since these polynomials obey the following recursive relation:
\begin{align}
    B_{n+1}(x_1,...,x_{n+1})=\sum_{i=0}^n\binom{n}{i}B_{n-i}(x_1,...,x_{n-i})x_{i+1},
\end{align}
one gets a recurrence relation for $p_n(t)$:
\begin{align}
    p_{n}(t)&=\frac{1}{n!}B_n\big(1!(\lambda\ast Q_1)(t),\dots,n!(\lambda\ast Q_n)(t)\big)\,p_0(t)\\&=\frac{1}{n}\sum_{k=0}^{n-1}\frac{k+1}{(n-(k+1))!}B_{n-(k+1)}\big(1!(\lambda\ast Q_1,...,\big(n-(k+1)\big)!\lambda\ast Q_{n-(k+1)}\big)(\lambda\ast Q_{k+1})(t)p_0(t)\\ &=\frac{1}{n}\sum_{k=1}^{n}kp_{n-k}(t)(\lambda\ast Q_{k})(t). \label{eq:pn-recursive}
\end{align}

\subsection{Homogeneous Poisson Train: Transient and Steady-State Probabilities}\label{subsec:hom-pos-train-steady-state}

For the Poisson train ($\lambda =\text{const}$), Eq.~(\ref{eq:p0}) specializes to
\begin{equation}
\begin{split}
     p_0(t) &=\exp{\Big( \lambda \int_0^t \text{d}s \;(Q_0(s) -1)\Big )}  \\
     &= \exp \left( \lambda \sum_{m=0}^{\infty} q_m \int_{0}^{t} ds \; ((1-e^{-s})^m -1)\right)\\
     &= \exp \left( \lambda \sum_{m=0}^{\infty} q_m \int_{e^{-t}}^{1} \frac{(1-u)^m-1}{u} du \right) \\
     &= \exp\left(-\lambda \sum_{m=0}^{\infty} q_m \left(H_m + \sum_{k=1}^{m} \binom{m}{k} \frac{(-1)^ke^{-kt}}{k}\right)\right),
\end{split}
\end{equation}
where $H_m = \sum_{k=1}^m \frac{1}{k}$ are the harmonic numbers.
Similarly, we may rewrite
\begin{equation}
\begin{split}
    \lambda\ast Q_k(t) &= \lambda \sum_{m=0}^{\infty} q_m \binom{m}{k} \int_{0}^{t} ds \; e^{-sk} (1-e^{-s})^{m-k} \\
    &= \lambda \sum_{m=0}^{\infty} q_m \binom{m}{k}  \int_{e^{-t}}^{1} u^{k-1} (1-u)^{m-k} du \\
    &= \lambda \sum_{m=0}^{\infty} q_m \binom{m}{k} \left(\frac{(k-1)!(m-k)!}{m!} - 
 \sum_{j=0}^{m-k} \binom{m-k}{j}\frac{(-1)^j}{k+j}\,e^{-(k+j)t}\right) \\
 &= \lambda \sum_{m=k}^{\infty}\frac{q_m}{k} - 
\lambda \sum_{m=0}^{\infty} q_m \binom{m}{k}\sum_{j=0}^{m-k} \binom{m-k}{j}\frac{(-1)^j}{k+j}\,e^{-(k+j)t}.
\end{split}
\end{equation}
Plugging this into Eq.~(\ref{eq:pn-recursive}), we get the recursive relation
\begin{equation}
 p_n(t)=\frac{\lambda}{n}\sum_{k=1}^{n}p_{n-k}(t)  \left(\sum_{m=k}^{\infty}q_m - k
 \sum_{m=0}^{\infty} q_m \binom{m}{k}\sum_{j=0}^{m-k} \binom{m-k}{j}\frac{(-1)^j}{k+j}\,e^{-(k+j)t} \right).
\end{equation}
In the steady-state limit $t \to \infty$, we have 
\begin{equation}
\begin{split}
   \pi_0^{(P)} &:= p_0(t\to\infty)=\exp(-\lambda \sum_{m=1}^{\infty}q_mH_m), \\
   \pi_n^{(P)} &:= p_n(t\to\infty)= \frac{\lambda}{n} \sum_{k=0}^{n-1} \pi_k \sum_{l=n-k}^{\infty} q_l,
\end{split}
\end{equation}
as given in the main text. The cumulants can be determined from the generating function as follows:
\begin{equation}
\begin{split}
    K(s):=\ln G(z\!=\!e^s) &= \lambda \sum_{m=0}^{\infty } q_m \int_0^{\infty}  ((e^{-\tau}(e^s-1)+1)^m -1)\;d\tau \\
    &=\lambda \sum_{m=1}^{\infty } q_m \sum_{k=1}^m \binom{m}{k} \frac{(e^s-1)^k}{k} \\
    &=\lambda \sum_{m=1}^{\infty } q_m \sum_{k=1}^m \frac{e^{sk}-1}{k} \\
    &= \sum_{l=1}^{\infty} \frac{s^l}{l!} \underbrace{\lambda \sum_{m=1}^{\infty} q_m \sum_{k=1}^{m} k^{l-1}}_{\kappa_{\pi,l}} \label{eq:pois-train-cumulants}
\end{split}
\end{equation}

\subsection{Gamma-Interval Train: Post-Release Distribution}\label{sec:gamma-post-release}

As a generalization to both the homogeneous Poisson train and the fixed-interval train, we assume that the time intervals between releases follow a gamma distribution with mean $1/\lambda$ and shape parameter $\alpha$. The corresponding time density reads
\begin{equation}
    f(\tau)=\frac{(\lambda \alpha)^{\alpha}\tau^{\alpha-1}e^{-\lambda \alpha \tau}}{\Gamma (\alpha)}.
\end{equation}
For $\alpha =1$, this turns into an exponential distribution, corresponding to the homogeneous Poisson train. In the limit $\alpha \to \infty$, $f(t)$ approaches a $\delta$-distribution centered at $1/\lambda$, corresponding to the fixed-interval train. By allowing for other values of $\alpha$, our results generalize to settings where action potentials still occur randomly, but in a bunched ($\alpha < 1$) or anti-bunched ($\alpha > 1$) manner. 

It will prove convenient to express the results using the factorial moments
\begin{equation}
    \overline{\mu_k} := \mathbb{E}[X(X-1)\ldots (X-k+1)], \qquad \overline{\mu_0} := 1.
\end{equation}
From those, the non-central moments $\mu_k = \mathbb{E}[X^k]$ can be recovered as 
\begin{equation}
    \mu_k = \sum_{j=1}^k S(k,j) \overline{\mu_j}
\end{equation}
where $S(k,j)$ are the Stirling numbers of the second kind. Usually the moments $\mu_{q,k}$ of the release size distribution $q$ scale with the $k$-th power of the molecule count (true in the case of a normal distribution, or for a normal-binomial mixture) and the molecule count is quite high. Under these circumstances, the leading order dominates, and one may often approximate $\overline{\mu_k} \approx \mu_k$.

\subsubsection{Conditional Post-Release Distribution}

We start out by considering the pre-release distribution $p_{n|n_0}^- =\left\langle \text{Binom}(n|n_0,e^{-\tau})\right\rangle_{\tau\sim f(t)}$ conditioned on last post-release count $n_0$. For the fixed-interval case $\alpha \to \infty$, we simply have $p_{n|n_0}^{-(F)} = \text{Binom}(n|n_0,e^{-1/\lambda})$. For the Poisson train ($\alpha=1$), it was shown in the main text that $p_{n|n_0}^{-(P)} = \text{BetaBinom}(n|n_0,\lambda,1)$. Other possible values of $\alpha$ may be analyzed using a generating function approach. Using the probability generating function of the binomial distribution, the corresponding generating function $G^-_{|n_0}(z):=\sum_n p^-_{n | n_0}z^n$ calculates as
\begin{equation}
\begin{split}
    G^-_{|n_0}(z) &= \int_0^{\infty} d\tau \;f(\tau) \left(1 + (z-1)e^{-\tau }\right)^{n_0} \\
    &= \frac{(\lambda \alpha)^{\alpha}}{\Gamma(\alpha)} \int_0^{\infty} d\tau \; \tau^{\alpha-1}e^{-\lambda \alpha \tau}\left(1 + (z-1)e^{-\tau }\right)^{n_0} \\
    &=  \frac{(\lambda \alpha)^{\alpha}}{\Gamma(\alpha)} \sum_{k=0}^{n_0} \binom{n_0}{k} (z-1)^k \underbrace{\int_0^{\infty} d\tau \; \tau^{\alpha-1}e^{-(\lambda \alpha + k) \tau}}_{=\Gamma(\alpha)/(\lambda \alpha + k)^{\alpha}} \\
    &= \sum_{k=0}^{n_0} \underbrace{\frac{n_0!}{(n_0-k)!} \left( \frac{\lambda \alpha}{\lambda \alpha+k} \right)^{\alpha}}_{=\; \overline{\mu_{-|n_0,k}}} \frac{(z-1)^k}{k!}, \label{eq:Gminusn0}
\end{split}
\end{equation}
where we identify the factorial moments $\overline{\mu_{-|n_0,k}} = \frac{d^k}{dz^k} G^-_{n_0}(z)|_{z=1}$. In particular:
\begin{align}
    \mu_{-|n_0} &= \overline{\mu_{-|n_0,1}} = n_0 \left(\frac{\lambda \alpha}{\lambda \alpha + 1} \right)^{\alpha} \\
    \sigma^2_{-|n_0} &=\overline{\mu_{-|n_0,2}} + \mu_{-|n_0} -  \mu_{-|n_0}^2 = n_0 \left( (n_0-1)\left(\frac{\lambda \alpha}{\lambda \alpha +2} \right)^{\alpha} + \left(\frac{\lambda \alpha}{\lambda \alpha +1} \right)^{\alpha} - n_0 \left(\frac{\lambda \alpha}{\lambda \alpha +1} \right)^{2\alpha}\right)
\end{align}

The conditional post-release distribution $p^+_{n|n_0} = \sum_m p^-_{n-m|n_0}q_m $ follows from a discrete convolution with the release size distribution $q_m$. This implies that the probability generating functions of both distributions multiply,
\begin{equation}
    \begin{split}
     G^+_{|n_0}(z) &= G^q (z) G^-_{|n_0}(z) \\
     &= \left(\sum_{j=0}^{\infty}  \overline{\mu_{q,j}} \frac{(z-1)^j}{j!}\right) \left(\sum_{k=0}^{n_0} \overline{\mu_{-|n_0,k}} \frac{(z-1)^k}{k!}\right) \\
     &= \sum_{j=0}^{\infty}  \underbrace{\sum_{k=0}^j \binom{j}{k}\overline{\mu_{q,j-k}} \;\overline{\mu_{-|n_0,k}}}_{=\; \overline{\mu_{+|n_0,j}}}  \frac{(z-1)^j}{j!},\label{eq:Gplusn0}
    \end{split}
\end{equation}
and that the cumulants add,
\begin{align}
    \mu_{+|n_0} &= \langle m \rangle + \mu_{-|n_0}  = \langle m \rangle + n_0 \left(\frac{\lambda \alpha}{\lambda \alpha + 1} \right)^{\alpha},\\
    \sigma_{+|n_0}^2 &= \sigma_m^2+ \sigma_{-|n_0}^2 = \sigma_m^2+ n_0 \left( (n_0-1)\left(\frac{\lambda \alpha}{\lambda \alpha +2} \right)^{\alpha} + \left(\frac{\lambda \alpha}{\lambda \alpha +1} \right)^{\alpha} - n_0 \left(\frac{\lambda \alpha}{\lambda \alpha +1} \right)^{2\alpha}\right).
\end{align}
In the Poisson case $\alpha = 1$, we have
\begin{align}
    \mu_{+|n_0}^{(P)} &=  \langle m \rangle + n_0 \frac{\lambda }{\lambda + 1}, \\
    {\sigma_{+|n_0}^{2(P)}} &=  \sigma_m^2  + n_0\frac{\left(\frac{n_0}{\lambda +1} +1\right)\lambda}{(\lambda+1)(\lambda+2)}.
\end{align}
In the fixed-interval limit $\alpha \to \infty$, we use that $\lim\limits_{\alpha\to \infty}\left(\frac{\lambda \alpha}{\lambda \alpha+k} \right)^{\alpha} = \exp(-k/\lambda)$ and find 
\begin{align}
    \mu_{+|n_0}^{(F)} &= \langle m \rangle + n_0e^{-1/\lambda},   \\
{\sigma_{+|n_0}^{2(F)}} &= \sigma_m^2+n_0e^{-1/\lambda}(1-e^{-1/\lambda}).
\end{align}

\subsubsection{Steady-State Post-Release Distribution}

The conditional post-release probabilities $p^+_{n|n_0}$ derived above can be interpreted as transition probabilities of a stochastic process. To be precise, this process is a discrete-time Markov chain, where the state space is the post-release molecule count and each time-step corresponds to a release event. After many releases, this Markov chain settles into a steady state with occupation probabilities $\pi^+_n$ that obey the following self-consistency relation:
\begin{equation}
    \pi^+_n =\sum_{j=0}^{\infty} p^+_{n|j} \pi^+_j  \label{eq:post-release-self-consistency}
\end{equation}

Using the previous results from Eqs. (\ref{eq:Gminusn0}) and (\ref{eq:Gplusn0}), this can be translated into an equation for the generating function:
\begin{equation}
    \begin{split}
    G^+ (z) &=  \sum_{l=0}^{\infty} G^+_{|l}(z)  \pi^+_l \\
    &= \sum_{l=0}^{\infty}\sum_{j=0}^{\infty} \sum_{k=0}^j \binom{j}{k}\overline{\mu_{q,j-k}} \;\overline{\mu_{-|l,k}} \frac{(z-1)^j}{j!} \pi^+_l \\
    &= \sum_{j=0}^{\infty} \frac{(z-1)^j}{j!}\sum_{k=0}^j \binom{j}{k}\overline{\mu_{q,j-k}} \; \left( \frac{\lambda \alpha}{\lambda \alpha+k} \right)^{\alpha}  \sum_{l=k}^{\infty} \frac{l!}{(l-k)!}\pi^+_l \\
    &= \sum_{j=0}^{\infty} \frac{(z-1)^j}{j!}\sum_{k=0}^j \binom{j}{k}\overline{\mu_{q,j-k}} \; \left( \frac{\lambda \alpha}{\lambda \alpha+k} \right)^{\alpha}  \overline{\mu_{+,k}}
    \end{split}
\end{equation}
By expanding $G^+(z)$ into its factorial moments and comparing both sides of the equation, we can extract a recursive relation for the factorial moments:
\begin{equation}
    \overline{\mu_{+,j}} =  \frac{1}{1 - (\frac{\lambda \alpha}{\lambda \alpha +j})^{\alpha}}
    \sum_{k=0}^{j-1} \binom{j}{k}\overline{\mu_{q,j-k}} \; \left( \frac{\lambda \alpha}{\lambda \alpha+k} \right)^{\alpha}  \overline{\mu_{+,k}} \label{eq:post-release-factorial}
\end{equation}
In particular, 
\begin{align}
    \mu_+ &= \overline{\mu_{+,1}} =  \frac{\langle m \rangle}{1 - (\frac{\lambda \alpha}{\lambda \alpha +1})^{\alpha}}, \\
    \begin{split}
        \sigma_+^2 &=  \overline{\mu_{+,2}} + \mu_+ - \mu_+^2 = \frac{\langle m^2\rangle - \langle m \rangle + 2\langle m \rangle \left( \frac{\lambda \alpha}{\lambda \alpha +1}\right)^{\alpha} \mu_+}{1 - (\frac{\lambda \alpha}{\lambda \alpha +2})^{\alpha}} + \mu_+ - \mu_+^2 \\
        &= \frac{1}{1 - (\frac{\lambda \alpha}{\lambda \alpha +2})^{\alpha}} \left( \sigma_m^2+ \frac{1}{1-(\frac{\lambda \alpha}{\lambda \alpha +1})^{\alpha}}  \left( \frac{(\frac{\lambda \alpha}{\lambda \alpha +2})^{\alpha} - (\frac{\lambda \alpha}{\lambda \alpha +1})^{2\alpha}}{1-(\frac{\lambda \alpha}{\lambda \alpha +1})^{\alpha}}\langle m\rangle^2-((\tfrac{\lambda \alpha}{\lambda \alpha +2})^{\alpha} - (\tfrac{\lambda \alpha}{\lambda \alpha +1})^{\alpha})\langle m\rangle\right)  \right) 
    \end{split}
\end{align}
In the Poisson case $\alpha \to 1$, this simplifies to :
\begin{align}
     \mu^{(P)}_+ &= (\lambda +1)     \langle m \rangle, \label{eq:post-release-P-mean} \\ 
    {\sigma_+^2}^{(P)} &=  \frac{\lambda}{2} ( \langle m \rangle + \langle m^2 \rangle)+ \sigma_m^2 
\end{align}
In the fixed-interval case $\alpha \to \infty$, again using $\lim\limits_{\alpha\to \infty}\left(\frac{\lambda \alpha}{\lambda \alpha+k} \right)^{\alpha} = \exp(-k/\lambda)$:
\begin{align}
    \mu^{(F)}_+ &= \frac{\langle m \rangle}{1-e^{-1/\lambda}},  \label{eq:post-release-F-mean}\\ 
    {\sigma_{+}^2}^{(F)} &=  \frac{\sigma_m^2 + e^{-1/\lambda} \langle m \rangle}{1-e^{-2/\lambda}}. \label{eq:post-release-F-var}
\end{align}

\subsection{\label{sec:appendix-fixed-train} Fixed-Interval Train: Steady State }

%In Fig. 4B, we compare the two post-release size distributions (left inset), and also plot the steady state distribution some time $T$ after the release.

\new{For the fixed-interval train, there is no steady state in the conventional sense, since the release events are fixed and the molecule count therefore always depends on the time  since the last release event $T = t - \lfloor t \lambda \rfloor / \lambda$ ($0 \le T \le 1/ \lambda$). To compare to the steady-state distribution $\pi^{(P)}$ of the time-homogeneous Poisson train (see Section~\ref{subsec:hom-pos-train-steady-state}), we define $\pi^{(F)}(T)$ as the distribution obtained from the steady-state post-release distribution $\pi^{+(F)}$  after a degradation time $T$.}

%\begin{align}
%\mu_{\pi}^{(F)}(T) &= \frac{e^{-T}}{1-e^{-1/\lambda}} \langle m \rangle, \\
%        {\sigma_{\pi}^2}^{(F)} (T) &=  \mu_{\pi}^{(F)}(T) + \frac{e^{-2T}}{1-e^{-2/\lambda}}(\sigma_m^2 - \langle m \rangle).   \label{eq:stat-f-mean-var}
%\end{align}

Again, we use the fact that the number of molecules remaining at time $T$ after a release with post-release count $n_0$ is binomially distributed with $N=n_0$ and $p=e^{-T}$, which we write as $\mathcal{B}^{n_0}(T)$. The equation for $\pi^{(F)}(T)$ therefore reads
\begin{equation}
    \pi_n^{(F)}(T) = \sum_{k=0}^{\infty} \mathcal{B}_n^{k}(T)\; \pi_k^{+(F)}. \label{eq:pi-F-T-general}
\end{equation}
Using the binomial generating function 
\begin{equation}
    G_{\mathcal{B}^{n_0}(T)}(z) = \sum_{n=0}^{n_0} \mathcal{B}_n^{n_0}(T) z^n=\sum_{n=0}^{n_0} \frac{n_0!}{(n_0-n)!} e^{-nT} \frac{(z-1)^n}{n!},
\end{equation}
the probability generating function becomes
\begin{equation}
G^{(F)}_T (z) = \sum_{k=0}^{\infty} G_{\mathcal{B}^{k}(T)}(z)\; \pi_k^{+(F)} = \sum_{j=0}^{\infty} \frac{(z-1)^j}{j!} e^{-jT}\sum_{k=j}^{\infty} \frac{k!}{(k-j)!}\pi^{+(F)}_k = \sum_{j=0}^{\infty} \frac{(z-1)^j}{j!} e^{-jT}  \;\overline{\mu_{+,j}}^{(F)}.
\end{equation}
Identifying the coefficients, we find the following relation between the factorial moments of the steady-state and the post-release distribution:
\begin{equation}
    \overline{\mu_{j}}^{(F)} (T)=e^{-jT}  \;\overline{\mu_{+,j}}^{(F)} \;\overset{(\ref{eq:post-release-factorial})}{=} \; \frac{e^{-jT}}{1 - e^{-j/\lambda}}
    \sum_{k=0}^{j-1} \binom{j}{k} \overline{\mu_{q,j-k}} \; \overline{\mu_{+,k}}^{(F)} \;e^{-k/\lambda}
\end{equation}
In particular, using the previous results from Eqs. (\ref{eq:post-release-F-mean}) and (\ref{eq:post-release-F-var}), the steady-state mean and variance calculate as
\begin{align}
        \mu^{(F)}(T) &= \overline{\mu_{1}}^{(F)} (T)=e^{-T} \mu_+^{(F)} =\frac{\langle m \rangle e^{-T}}{1-e^{-1/\lambda}} \\
        \begin{split}
        {\sigma^2}^{(F)} (T) &= \overline{\mu_{2}}^{(F)} (T)- (\mu^{(F)}(T))^2  + \mu^{(F)}(T)  \\
            &= e^{-2T} (\sigma_+^{2(F)}+ \mu_+^{2(F)} - \mu_+^{(F)}) - e^{-2T}\mu_+^{2(F)}  + \mu^{(F)}(T)  \\
            &= e^{-2T} (\sigma_+^{2(F)} - \mu_+^{(F)}) + \mu^{(F)}(T)\\
            &= \frac{e^{-2T}(\sigma_m^2 - \langle m \rangle) }{1-e^{-2/\lambda}}+ \mu_{\pi}^{(F)}(T) 
        \end{split} \label{eq:sigma2-F-T}
\end{align}
For comparison, we read from Eq. (\ref{eq:pois-train-cumulants}) for the steady state mean and variance of the Poisson train:
\begin{align}
    \mu^{(P)} &= \lambda    \langle m \rangle, \label{eq:mu-P} \\ 
    {\sigma^2}^{(P)} &=  \frac{\lambda}{2} ( \langle m \rangle + \langle m^2 \rangle) \label{eq:sigma-P}
\end{align}

\begin{figure*}[th]
    \centering
    \includegraphics[width=\textwidth]{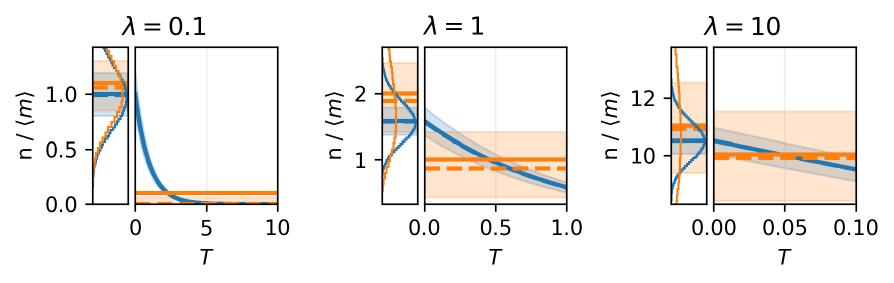}
    \caption{\new{Post-release distribution $\pi^+$ (left) and quartiles of the steady state distribution $\pi$ (right) for the fixed-interval train (blue) and the Poisson train (orange) and different release rates $\lambda$. The steady-state distribution of the fixed train depends on the time $T$ since the last release event. As in Figure 4 in the main text, the count of released vesicles is assumed to follow a binomial distribution with $V=50$ release sites and release probability $\rho=0.2$. The quantal size distribution is normal with mean $\langle c \rangle=1000$ molecules per vesicle and $\sigma_c=0.2\langle c \rangle$. The quartiles of the distributions were obtained from stochastic sampling (see III) and are displayed in the same scheme as in Figure 4 of the main text. }}
    \label{fig:steady-state-comparison}
\end{figure*}

\new{In Fig.~\ref{fig:steady-state-comparison}, we compare both the post-release distributions (left inset) and steady state distributions (right inset) of the fixed-interval and Poisson train. Assuming equal release sizes, $\mu^{(F)}(T)$ (continuous blue line) crosses $\mu^{(P)}$ (continuous orange line) in the first half of the interval $1/\lambda$. This means that for most of the time, the mean molecule count from the fixed train lies below the mean molecule count from the Poisson train. Note that this is not necessarily the case for the median (dashed line in Fig.~\ref{fig:steady-state-comparison}), especially for low $\lambda$. Due to the stochastic release timing, the distribution for the Poisson train appears wider and more asymmetric than the fixed train. From the above equations, one can show that for $\lambda > 2 \frac{\sigma_m^2}{\langle m \rangle ^2}$, the variance ${\sigma_{\pi}^2}^{(P)}$ is always strictly greater than  ${\sigma_{\pi}^2}^{(F)}(T)$.}

\subsection{Fixed-Interval Train: Normal Approximation}

The plots \new{in Fig.~\ref{fig:steady-state-comparison}} indicate that the steady-state distribution for the fixed-interval train may often be approximated by a normal distribution. Here, we provide some mathematical intuition for this suggestion. We base our derivation on three assumptions:
\begin{enumerate}
    \item The release size distribution is approximately normal. 
    \item The variance of the release size is much greater than the mean molecule count, $\langle m \rangle  / \sigma_m^2  \ll (1-e^{-1/\lambda})$.
    \item The pre-release distribution $p_{n|n_0}^{-(F)} = \text{Binom}(n|n_0,e^{-1/\lambda})$ is approximately normal.
\end{enumerate}

Provided the vesicles follow binomial release statistics (see Section IID in the main text), assumption 1 is justified if the release probability $\rho$ is not too close to $0$ or $1$ and the  number of releasable vesicles is $V \gtrsim  20$. It can also hold for $V \lesssim 20 $ if the variance in vesicle content $\sigma_c^{2}$ is high. Assumption 2 holds if the mean content $\langle c \rangle$ is not too small, and if $\lambda$ is not too high. Assumption 3 follows from Assumption 1, provided $\lambda$ is not too small.

Using that the post-release count is the sum of the pre-release count and the release size, and that the sum of normal variables is again normal, we may then approximate 
\begin{equation}
    p^{+(F)}(x|y) \approx \mathcal{N}(x\;|\;e^{-1/\lambda} y + \langle m\rangle,  ye^{-1/\lambda}(1-e^{-1/\lambda})+ \sigma_m^2) \approx \mathcal{N}(x\;|\;e^{-1/\lambda} y + \langle m\rangle,  \sigma_m^2)
\end{equation}
where we eliminated the dependence on $y$ in the variance, using Assumption 2. The self-consistency relation Eq.~(\ref{eq:post-release-self-consistency}) becomes 
\begin{equation}
    \pi^{+(F)}(x) \approx \int dy \;\mathcal{N}(x\;|\;e^{-1/\lambda} y + \langle m\rangle,  \sigma_m^2) \; \pi^{+(F)}(y).
\end{equation}
Using a normal distribution as an ansatz for $\pi^{+(F)}(x)$ and performing the Gaussian integral  yields
\begin{equation}
    \pi^{+(F)}(x) \approx \mathcal{N}\left(x\,\mid\,\frac{\langle m \rangle}{1-e^{-1/\lambda}},  \frac{\sigma_m^2}{1-e^{-2/\lambda}}\right).
\end{equation}
Similarly, Eq.~(\ref{eq:pi-F-T-general}) becomes
\begin{equation}
\begin{split}
     \pi^{(F)}(x,T) &\approx \int dy \; \delta(x-ye^{-T}) \pi^{+(F)}(y) \\
     &= \mathcal{N}\left(x\,\mid\,\frac{\langle m \rangle e^{-T}}{1-e^{-1/\lambda}},  \frac{\sigma_m^2e^{-2T}}{1-e^{-2/\lambda}}\right)
\end{split}
\end{equation}
which corresponds to the mean and variance in Eq.~(\ref{eq:sigma2-F-T}), again only accounting for  terms of highest order in the quantal size $\langle c \rangle$, as implied by Assumption 2.

\section{Extended Model with Vesicle Depletion}\label{sec:appendix-extended-model}

\subsection{Release Size Distribution}

As a prerequisite for the following sections, we investigate the properties of the release size distribution
\begin{equation}
     q_{m,v|u}=\left(q^{(C)}\right)^{*v}_m \text{Binom}(v \;|\;V-u, \rho),
\end{equation}
where $m$ is the count of released molecules, $v$ is the count of released vesicles, $q^{(C)}$ is the quantal size distribution with mean $\langle c \rangle$ and variance $\sigma_c^2$, $(\cdot)^{*v}$ is the $v$-fold convolution, $V$ is the total number of vesicle docking sites, $u$ is the number of empty docking sites and $\rho$ is the release probability per docked vesicle. 

We write the bivariate generating function as
\begin{equation}
\begin{split}
    G^q_{|u}(z,s) &= \sum_{m=0}^{\infty}  \sum_{v=0}^{\infty} z^m s^v\left(q^{(C)}\right)^{*v}_m \text{Binom}(v \;|\;V-u, \rho) \\
    &=   \sum_{v=0}^{\infty} s^v (G^{q(C)}(z))^v\; \text{Binom}(v \;|\;V-u, \rho) \\
    &= (1-\rho + \rho s G^{q(C)}(z))^{V-u} \label{eq:release-size-gen-func}
\end{split}
\end{equation}
Equivalently, the cumulant generating function $K^q(z,s)$ is given as
\begin{equation}
    K^q(z,s) =\ln G^q(e^z, e^s) = (V-u) \ln(1-\rho+\rho \exp(s+K^{q(C)}(z)))
\end{equation}
and the first few cumulants can be computed using the total derivatives:
\begin{align}
    \kappa_{q|u,(1,0)} &= \frac{d}{dz}K^q(z,s)|_{0,0} = (V-u)\rho \langle c\rangle \label{eq:depl-gamma-size-mean-m}\\
    \kappa_{q|u,(0,1)} &= \frac{d}{ds}K^q(z,s)|_{0,0} = (V-u)\rho\\
    \kappa_{q|u,(2,0)} &= \frac{d^2}{dz^2}K^q(z,s)|_{0,0} = (V-u)\rho (\sigma_c^2+(1-\rho)\langle c\rangle^2) \\
    \kappa_{q|u,(1,1)} &=\frac{d^2}{dz\;ds}K^q(z,s)|_{0,0} = (V-u)(1-\rho)\rho \langle c \rangle\\
    \kappa_{q|u,(0,2)} &= \frac{d^2}{ds^2}K^q(z,s)|_{0,0} = (V-u)(1-\rho)\rho \label{eq:depl-gamma-size-var-v}
\end{align}

\subsection{General Poisson Train: Mean Molecule Count}

\subsubsection{Transient Mean}

Unlike for the Master Equation of the simpler model discussed in Section~\ref{sec:appendix-derivation-probability-density}, explicitly solving the Master Equation
\begin{equation}
    \dt{p_{n,u}}= -( n+\lambda+ \xi u)p_{n,u}+ (n+1)p_{n+1,u} + \xi (u+1)p_{n,u+1} + \lambda\sum_{m,v} q_{m,v|u-v} p_{n-m,u-v}
\end{equation}
via generating functions does not seem feasible, due to the additional dependence of the release size on the current count of empty release sites. Still, by taking expectations, one can derive coupled ODEs for the moments, as demonstrated by Gambrell et al. \cite{gambrell2024feedforward}. The equations for the means read:
\begin{align}
    \frac{d\langle u \rangle}{dt} &= (V - \langle u\rangle) \rho \lambda(t) - \xi \langle u\rangle, \\
    \frac{d\langle n \rangle}{dt} &= (V - \langle u\rangle) \langle c\rangle \rho\lambda(t)  - \langle n \rangle
\end{align}
Solving this for the initial conditions $\langle u(0)\rangle = u_0$ and $\langle n(0)\rangle =n_0$, one finds
\begin{align}
\langle u (t)\rangle &= u_0 e^{-(\xi + \rho \lambda) t}+  \frac{V \rho \lambda}{\xi + \rho \lambda } (1-e^{-(\xi + \rho \lambda) t}),  \\
\langle n (t) \rangle&= n_0e^{-t} + \langle c \rangle \rho \lambda \left[\frac{V\xi}{\xi + \rho \lambda}(1-e^{-t }) -   \left(u_0 - \frac{\rho V \lambda}{\xi + \rho\lambda}\right) \frac{e^{-\xi t}-e^{-t} }{1-\xi} \right]
\end{align}
for the homogeneous case ($\lambda = \text{const}$) and 
\begin{align}
\langle u (t)\rangle &= 
e^{-\xi t - \rho \int_0^t \lambda(r)\,dr} 
\left( 
u_0 + V \rho \int_0^t e^{\xi s + \rho \int_0^s \lambda(r)\,dr} \, \lambda(s)\, ds 
\right), \\
\langle n (t) \rangle&= n_0e^{-t}  + \langle c \rangle \rho \int_0^t e^{s-t} (V - \langle u (s)\rangle) \lambda(s)\, ds
\end{align}
for the general case $\lambda= \lambda(t)$. In particular, we may express the correction to the mean molecule count without depletion $\langle n(t)|u=0\rangle$ as:
\begin{equation}
\begin{split}
    \langle \Delta n(t) \rangle &= \langle n(t)|u=0\rangle  - \langle n(t)\rangle = \langle c \rangle \rho \int_0^t e^{s-t} \langle u(s)  \rangle\lambda(s)\, ds 
    \\ &= \langle c \rangle \rho e^{-t}  \int_0^t e^{-(\xi-1) s - \rho \int_0^s \lambda(r)\,dr} \left( u_0 + V \rho \  \int_0^s e^{\xi \tau}e^{\rho \int_0^\tau \lambda(r)\,dr} \, \lambda(\tau)\, d\tau \right) \lambda(s)\, ds 
\end{split}
\end{equation}
Introducing  $\lambda_{\text{min}} := \text{min}_{\tau \le t}[\lambda(t)], \lambda_{\text{max}} := \text{max}_{\tau \le t}[\lambda(t)]$ and using the identities
\begin{align}
    \frac{\;\lambda_{\text{min}} \Big(\exp(at+\rho\lambda_{\text{min}} t)-1 \Big)}{a+\rho\;\lambda_{\text{min}} }  \le \int_0^t e^{as} e^{+ \rho \int_0^s \lambda(r)dr} \lambda(s) \;ds \le \frac{\lambda_{\text{max}, t}\Big(\exp(at+\rho\lambda_{\text{max}} t)-1\Big)  }{a+\rho\;\lambda_{\text{max}} }, \\ 
    \frac{\lambda_{\text{min}, t}\Big(\exp(at-\rho\lambda_{\text{max}} t)-1\Big) }{a-\rho\;\lambda_{\text{max}} } \le \int_0^t e^{as} e^{- \rho \int_0^s \lambda(r)dr} \lambda(s) \;ds \le \frac{\lambda_{\text{max}}  \Big(\exp(at-\rho\lambda_{\text{min}} t)-1\Big)}{a-\rho\;\lambda_{\text{min}} }, \\ 
\end{align}
we find
\begin{align}
\langle \Delta n(t) \rangle \; &\ge \;\langle c\rangle \rho\, \lambda_{\min}\, e^{-t} \left[ \left(u_0 - \frac{\rho V \lambda_{\max}}{\xi + \rho\lambda_{\min}}\right) \frac{e^{(1-\xi-\rho(\lambda_{\min}-\lambda_{\max}))t}-1}{1-\xi-\rho(\lambda_{\min}-\lambda_{\max})} + \frac{\rho V \lambda_{\min}}{\xi + \rho\lambda_{\min}}\, \frac{e^{(1+\rho(\lambda_{\min}-\lambda_{\max}))t}-1}{1+\rho(\lambda_{\min}-\lambda_{\max})} \right] \\
    \langle \Delta n(t) \rangle \; &\le \; \langle c\rangle \rho\, \lambda_{\max}\, e^{-t} \left[ \left(u_0 - \frac{\rho V \lambda_{\min}}{\xi + \rho\lambda_{\max}}\right) \frac{e^{(1-\xi-\rho(\lambda_{\max}-\lambda_{\min}))t}-1}{1-\xi-\rho(\lambda_{\max}-\lambda_{\min})} + \frac{\rho V \lambda_{\max}}{\xi + \rho\lambda_{\max}}\, \frac{e^{(1+\rho(\lambda_{\max}-\lambda_{\min}))t}-1}{1+\rho(\lambda_{\max}-\lambda_{\min})} \right].
\end{align}
We see that the correction is linear in $u_0$ and grows with $O(1/\xi)$ as $\xi \to \infty$.

\subsubsection{Post-Release Mean}

The pre-release distribution, given that the last event occurred a time $T$ ago and resulted in $n_0$ molecules and $u_0$ empty release sites, is given by:
\begin{equation}
\begin{split}
    p^-_{n,u|n_0,u_0}(t,T) = \langle \mathcal{B}^{n_0}_n(T+\tau) \mathcal{B}^{u_0}_u(\xi(T+\tau))\rangle_{\tau \sim \text{Exp}(\lambda(t))}, \label{eq:depl-cond-pre-release-general}
\end{split}
\end{equation}
where again $\mathcal{B}^{n_0}_n(t):=\text{Binom}(n|n_0,e^{-t})$. The probability of having $n$ molecules and $u$ empty release sites after the next release follows from the convolution with the release size distribution: 
\begin{equation}
    p_{n,u|n_0,u_0}^+(t,T) = \sum_{j,w}q_{n-j,u-w|w} p^-_{j,w|n_0,u_0}(t,T). \label{eq:conditional-hitting-prob}
\end{equation}
Taking the mean of this expression, one finds
\begin{equation}
    \mu_{+|n_0,u_0} (t,T) =\mu_{+|n_0,0} (t,T) - u_0 \langle c \rangle \rho  \langle e^{-\xi(T+\tau)} \rangle_{\tau \sim \text{Exp}(\lambda(t))},
\end{equation}
where 
\begin{equation}
    \mu_{+|n_0,0} (t,T) = \langle c \rangle\rho V+ n_0 \langle e^{-(T+\tau)}\rangle_{\tau \sim \text{Exp}(\lambda(t))}
\end{equation}
is the mean without vesicle depletion. Again, we see that the effect of vesicle depletion results in an additional term, which is proportional to $u_0 \langle e^{-\xi(T+\tau)} \rangle_{\tau \sim \text{Exp}(\lambda(t))}$. Considering the identity,
\begin{equation}
    \frac{\text{min}_{\tau \ge t}[\lambda(\tau)] }{\text{max}_{\tau \ge t}[\lambda(\tau)] + \xi } \le\langle e^{-\xi\tau} \rangle_{\tau \sim \text{Exp}(\lambda(t))} \le \frac{\text{max}_{\tau \ge t}[\lambda(\tau)] }{\text{min}_{\tau \ge t}[\lambda(\tau)] + \xi },
\end{equation}
we may again conclude that the correction has a leading order of $O(1/\xi)$ as $\xi \to \infty$.

\subsection{Gamma-Interval Train: Post-Release Distribution}

We now turn to the case of constant release rates. As in Section~\ref{sec:gamma-post-release}, we suppose that the time between releases is gamma distributed with mean $1/\lambda$ and shape-parameter $\alpha$, which may serve as a generalization of both the homogeneous Poisson train and the fixed-interval train. Analogously to  Eq.~(\ref{eq:depl-cond-pre-release-general}), the conditional pre-release distribution reads
\begin{equation}
    p^-_{n,u|n_0,u_0} = \langle \mathcal{B}^{n_0}_n(\tau) \mathcal{B}^{u_0}_u(\xi\tau)\rangle_{\tau \sim \text{Gamma}(\alpha,\alpha\lambda)},
\end{equation}
where $n_0$ and $u_0$ are the previous post-release count of molecules and empty vesicle sites, respectively.  Using the binomial generating function, we find for the bivariate generating function $G^-_{|n_0, u_0}(z,s):=\sum_n p^-_{n,u | n_0,u_0}z^ns^u$:
\begin{equation}
\begin{split}
    G^-_{|n_0,u_0}(z,s) &= \int_0^{\infty} d\tau \;f(\tau) \left(1 + (z-1)e^{-\tau }\right)^{n_0}\left(1 + (s-1)e^{-\xi\tau }\right)^{u_0} \\
    &= \frac{(\lambda \alpha)^{\alpha}}{\Gamma(\alpha)} \int_0^{\infty} d\tau \; \tau^{\alpha-1}e^{-\lambda \alpha \tau}\left(1 + (z-1)e^{-\tau }\right)^{n_0}\left(1 + (s-1)e^{-\xi\tau }\right)^{u_0} \\
    &=  \frac{(\lambda \alpha)^{\alpha}}{\Gamma(\alpha)} \sum_{k=0}^{n_0}\sum_{w=0}^{u_0} \binom{n_0}{k} \binom{u_0}{w} (z-1)^k (s-1)^w \underbrace{\int_0^{\infty} d\tau \; \tau^{\alpha-1}e^{-(\lambda \alpha + k+ w\xi) \tau}}_{=\Gamma(\alpha)/(\lambda \alpha + k + w\xi)^{\alpha}} \\
    &= \sum_{k=0}^{n_0}\sum_{w=0}^{u_0} \underbrace{\frac{n_0!}{(n_0-k)!} \frac{u_0!}{(u_0-w)!} \left( \frac{\lambda \alpha}{\lambda \alpha+k + w\xi} \right)^{\alpha}}_{=\; \overline{\mu_{-|n_0,u_0,(k,w)}}} \frac{(z-1)^k}{k!}\frac{(s-1)^w}{w!},  \label{eq:depl-gamma-cond-pre-release}
\end{split}
\end{equation}
where we identify the multivariate factorial moments $\overline{\mu_{-|n_0,u_0,(k,w)}} = \frac{d^k}{dz^k} \frac{d^w}{ds^w} G^-_{|n_0,u_0}(z,s)|_{z,s=(1,1)}$. For later use, we also compute the first few cumulants of this distribution. Introducing 
\begin{equation}
    C_{kw}:= \left( \frac{\lambda \alpha}{\lambda \alpha+k + w\xi} \right)^{\alpha},
\end{equation}
we find
\begin{align}
    \kappa_{-|n_0,u_0,(1,0)} &= \overline{\mu_{-|n_0,u_0,(1,0)}} =n_0C_{10}, \\
    \kappa_{-|n_0,u_0,(0,1)} &= \overline{\mu_{-|n_0,u_0,(0,1)}}=u_0 C_{01}, \\
    \kappa_{-|n_0,u_0,(2,0)} &= \overline{\mu_{-|n_0,u_0,(2,0)}} +  \overline{\mu_{-|n_0,u_0,(1,0)}}(1- \overline{\mu_{-|n_0,u_0,(1,0)}}) = n_0^2(C_{20}-C_{10}^2) + n_0(C_{10}-C_{20}), \\
    \kappa_{-|n_0,u_0,(1,1)} &= \overline{\mu_{-|n_0,u_0,(1,1)}} - \overline{\mu_{-|n_0,u_0,(1,0)}}\;\overline{\mu_{-|n_0,u_0,(0,1)}}= n_0u_0 (C_{11}- C_{10} C_{01}) ,\\
    \kappa_{-|n_0,u_0,(0,2)} &= \overline{\mu_{-|n_0,u_0,(0,2)}} +  \overline{\mu_{-|n_0,u_0,(0,1)}}(1- \overline{\mu_{-|n_0,u_0,(0,1)}}) = u_0^2(C_{02}-C_{01}^2) + u_0(C_{01}-C_{02}) .
\end{align}

\subsubsection{Conditional Post-Release Distribution}

Given the pre-release distribution $p^-$ and the release size distribution $q$, the post-release distribution follows as

\begin{equation}
    p_{n,u|n_0,u_0}^+ = \sum_{j,v}q_{n-j,u-v|v} p^-_{j,v|n_0,u_0}.
\end{equation}

Consequently, the probabilitiy generating function can be expressed as:

\begin{equation}
    \begin{split}
     G^+_{|n_0, u_0}(z,s) &= \sum_{n=0}^{\infty} \sum_{u=0}^{\infty} \sum_{j=0}^{\infty} \sum_{v=0}^{\infty} z^ns^u q_{n-j, u-v|v}p^-_{j,v|n_0,u_0} \\
     &=  \sum_{j=0}^{\infty} \sum_{v=0}^{\infty} z^j s^v \left(\sum_{n=0}^{\infty} \sum_{u=0}^{\infty} z^ns^u q_{n, u|v} \right) p^-_{j,v|n_0,u_0} \\
     &\overset{\ref{eq:release-size-gen-func}}{=} \sum_{j=0}^{\infty} \sum_{v=0}^{\infty} z^j s^v  (1-\rho + \rho s G^{q(C)}(z))^{V-v} p^-_{j,v|n_0,u_0} \\
     &= (1-\rho + \rho s G^{q(C)}(z))^{V} \sum_{j=0}^{\infty} \sum_{v=0}^{\infty} z^j s^v  (1-\rho + \rho s G^{q(C)}(z))^{-v} p^-_{j,v|n_0,u_0} \\
     &= G^q_{|0}(z,s) \underbrace{G^-_{|n_0,u_0} \left(z, \frac{s}{1-\rho + \rho s G^{q(C)}(z)}\right) }_{:= G^h_{|n_0,u_0}(z,s)} \label{eq:depl-gamma-cond-post-release}
    \end{split}
\end{equation}
For convenience, we introduce a new probability distribution $h_{|n_0,u_0}$ that is characterized by its generating function $G^h_{|n_0,u_0}(z,s)=G^-_{|n_0,u_0} \left(z, \frac{s}{1-\rho + \rho s G^{q(C)}(z)}\right)$. As suggested by the last line of Eq. (\ref{eq:depl-gamma-cond-post-release}), the conditional post-release count is the sum of the release size distribution under a full vesicle pool, and a variable following the distribution $h_{|n_0,u_0}$. In particular, the cumulants of both distributions add up. This suggest switching to the cumulant generating function $K(z,s) =\ln G(e^z, e^s)$. By evaluating the derivatives of 
\begin{equation}
    K^h_{|n_0,u_0} (z,s)= K^-_{|n_0,u_0}\Big(z,\;s-\ln(1-\rho+\rho \exp(s+K^{q(C)}(z))\Big) 
\end{equation}
at $(z,s)=(0,0)$,  we find the cumulants 
\begin{align}
    \kappa_{h|n_0,u_0,(1,0)} &= \frac{d}{dz}K^-_{|n_0,u_0}(z,s)|_{0,0} = \kappa_{-|n_0,u_0,(1,0)} - \rho \langle c \rangle \kappa_{-|n_0,u_0,(0,1)} \label{eq:depl-cond-cumu-start}, \\
    \kappa_{h|n_0,u_0,(0,1)} &= \frac{d}{ds}K^-_{|n_0,u_0}(z,s)|_{0,0} = (1-\rho) \kappa_{-|n_0,u_0,(0,1)}, \\
    \begin{split}
        \kappa_{h|n_0,u_0,(2,0)} &= \frac{d^2}{dz^2}K^-_{|n_0,u_0}(z,s)|_{0,0} \\
        &= (\rho(\rho-1)\langle c\rangle ^2+\sigma_c^2)\kappa_{-|n_0,u_0,(0,1)} + \rho^2 \langle c \rangle^2 \kappa_{-|n_0,u_0,(0,2)} -2\rho \langle c \rangle\kappa_{-|n_0,u_0,(1,1)} + \kappa_{-|n_0,u_0,(2,0)},
    \end{split}
    \\
    \kappa_{h|n_0,u_0,(1,1)} &=  \frac{d^2}{dz\,ds}K^-_{|n_0,u_0}(z,s)|_{0,0} =  (1-\rho) (\kappa_{-|n_0,u_0,(1,1)} - \rho \langle c \rangle (\kappa_{-|n_0,u_0,(0,1)} + \kappa_{-|n_0,u_0,(0,2)}) ), \\
    \kappa_{h|n_0,u_0,(0,2)} &= \frac{d^2}{ds^2}K^-_{|n_0,u_0}(z,s)|_{0,0} = (1-\rho)^2 \kappa_{-|n_0,u_0,(0,2)} - \rho (1-\rho)\kappa_{-|n_0,u_0,(0,1)} \label{eq:depl-cond-cumu-end}.
\end{align}

From these results, we evaluate the post-release mean and variance as: 
\begin{align}
\begin{split}
     \mu_{+|n_0,u_0} &= \kappa_{+|n_0,u_0,(1,0)} = \kappa_{q|0,(1,0)} + \kappa_{h|n_0,u_0,(1,0)} \\
     &= V\rho\langle c\rangle +  n_0C_{10} - \rho\langle c\rangle u_0C_{01},
\end{split}\\
\begin{split}
    \sigma^2_{+|n_0,u_0} &= \kappa_{+|n_0,u_0,(2,0)} = \kappa_{q|0,(2,0)} + \kappa_{h|n_0,u_0,(2,0)} \\
    &= V\rho(\sigma_c^2+(1-\rho)\langle c\rangle^2) + (\rho(\rho-1)\langle c\rangle ^2+\sigma_c^2)u_0C_{01} + \rho^2 \langle c \rangle^2 (u_0^2(C_{02}-C_{01}^2) + u_0(C_{01}-C_{02}) ) \\\
    &\quad-2\rho \langle c \rangle n_0u_0 (C_{11}- C_{10} C_{01}) + n_0^2(C_{20}-C_{10}^2) + n_0(C_{10}-C_{20}).
\end{split}
\end{align}

For the fixed train, we specialize $C_{kw} = e^{-(k+w\xi)/\lambda}$: 

\begin{align}
\begin{split}
     \mu_{+|n_0,u_0}^{(F)} &= V\rho\langle c\rangle +  n_0e^{-1/\lambda} - \rho\langle c\rangle u_0e^{-\xi/\lambda},
\end{split}\\
\begin{split}
    \sigma_{+|n_0,u_0}^{2(F)} &= V\rho(\sigma_c^2+(1-\rho)\langle c\rangle^2)  + n_0e^{-1/\lambda}(1-e^{-1/\lambda})+ u_0\left((\rho(2\rho-1)\langle c\rangle ^2+\sigma_c^2)e^{-\xi/\lambda}- \rho^2 \langle c \rangle^2 e^{-2\xi/\lambda} \right) 
\end{split}
\end{align}

For the Poisson train, we have $C_{kw} = \frac{\lambda}{\lambda + k+w\xi}$. The expression simplifies to:

\begin{align}
\begin{split}
     \mu_{+|n_0,u_0}^{(P)} &= V\rho\langle c\rangle +  n_0 \frac{\lambda}{\lambda+1} - \rho\langle c\rangle u_0 \frac{\lambda }{\lambda+\xi}
\end{split}\\
\begin{split}
    \sigma_{+|n_0,u_0}^{2(P)} &= V\rho(\sigma_c^2+(1-\rho)\langle c\rangle^2) + n_0 \frac{(\frac{n_0}{\lambda+1}+1)\lambda}{(\lambda+2)(\lambda+1)} - \frac{2\rho \langle c \rangle n_0u_0\lambda \xi}{(\lambda+\xi+1)(\lambda+\xi)(\lambda+1)}  \\\
    &\quad + u_0 \left(\rho^2 \langle c \rangle^2 \frac{(\frac{\xi u_0}{\lambda+\xi}+ 1 )\lambda\xi}{(\lambda+2\xi)(\lambda+\xi)} + \frac{(\rho(\rho-1)\langle c\rangle ^2+\sigma_c^2)\lambda}{\lambda+\xi} \right)
\end{split}
\end{align}

\subsubsection{Steady-State Post-Release Distribution}

Next, we turn to the post-release distribution in steady-state, which can be determined from the following self-consistency relation:
\begin{equation}
\begin{split}
    G^+(z,s) &= \sum_{n_0=0}^{\infty}\sum_{u_0=0}^{\infty} G^+_{|n_0, u_0}(z,s) \pi^+_{n_0,u_0} \\
    &= \sum_{n_0=0}^{\infty}\sum_{u_0=0}^{\infty} G^q_{|0}(z,s) \sum_{k=0}^{n_0}\sum_{w=0}^{u_0} \frac{n_0!}{(n_0-k)!} \frac{u_0!}{(u_0-w)!} \left( \frac{\lambda \alpha}{\lambda \alpha+k + w\xi} \right)^{\alpha} \frac{(z-1)^k}{k!}\frac{( \frac{s}{1-\rho + \rho s G^{q(C)}(z)}-1)^w}{w!} \pi^+_{n_0,u_0} \\
    &= G^q_{|0}(z,s)  \sum_{k=0}^{\infty}\sum_{w=0}^{\infty}  \left( \frac{\lambda \alpha}{\lambda \alpha+k + w\xi} \right)^{\alpha} \frac{(z-1)^k}{k!}\frac{( \frac{s}{1-\rho + \rho s G^{q(C)}(z)}-1)^w}{w!} \sum_{n_0=k}^{\infty}\sum_{u_0=w}^{\infty} \frac{n_0!}{(n_0-k)!} \frac{u_0!}{(u_0-w)!} \pi^+_{n_0,u_0} \\
&= G^q_{|0}(z,s)  \underbrace{\sum_{k=0}^{\infty}\sum_{w=0}^{\infty}  \left( \frac{\lambda \alpha}{\lambda \alpha+k + w\xi} \right)^{\alpha}\overline{\mu_{+,(k,w)}} \frac{(z-1)^k}{k!}\frac{( \frac{s}{1-\rho + \rho s G^{q(C)}(z)}-1)^w}{w!}\; }_{:=G^{H}(z,s/(1-\rho + \rho s G^{q(C)}(z)))\; := G^{h}(z,s) } ,
\end{split}
\end{equation}
where we again introduce a new generating function $G^h(z,s)$, characterizing a probability distribution $h$. Note that the factorial moments of the post-release distribution are still present on the right-hand side. This suggests that there is no closed formula for the moments and that they have to be evaluated recursively. In the following, we show this up to second order. 

As for the conditional distribution, we again switch to the cumulant generating function
\begin{equation}
    K^h (z,s)= K^H\Big(z,\;s-\ln(1-\rho+\rho \exp(s+K^{q(C)}(z))\Big),
\end{equation}
where the cumulants $\kappa_{H,(k,w)}$ calculate from the factorial moments $\overline{\mu_{H,(k,w)}} =C_{kw} \;\overline{\mu_{+,(k,w)}}$ as follows:
\begin{align}
    \kappa_{H,(1,0)} &= \overline{\mu_{H,(1,0)}} = C_{10} \overline{\mu_{+,(1,0)}} =C_{10} \kappa_{+,(1,0)}  \\
    \kappa_{H,(0,1)} &= \overline{\mu_{H,(0,1)}}= C_{01} \overline{\mu_{+,(0,1)}} =C_{01} \kappa_{+,(0,1)} \\
    \begin{split}
        \kappa_{H,(2,0)} &= \overline{\mu_{H,(2,0)}} +  \overline{\mu_{H,(1,0)}}(1- \overline{\mu_{H,(1,0)}}) \\ &= C_{20}\overline{\mu_{+,(2,0)}} +  C_{10}\overline{\mu_{+,(1,0)}}(1- C_{10 }\overline{\mu_{+,(1,0)}}) \\
        &= C_{20} \kappa_{+,(2,0)} + (C_{20} - C_{10}^2) \kappa_{+,(1,0)}^2 +  (C_{10}-C_{20}) \kappa_{+,(1,0)}  
    \end{split} \\
    \begin{split}
        \kappa_{H,(1,1)} &= \overline{\mu_{H,(1,1)}} - \overline{\mu_{H,(1,0)}}\;\overline{\mu_{H,(0,1)}} \\
        &= C_{11} \overline{\mu_{+,(1,1)}} - C_{10} C_{01}\overline{\mu_{+,(1,0)}}\;\overline{\mu_{+,(0,1)}} \\
        &= C_{11}\kappa_{+,(1,1)} + (C_{11}- C_{10} C_{01}) \kappa_{+,(1,0)} \kappa_{+,(0,1)} 
    \end{split} \\
    \begin{split}
        \kappa_{H,(0,2)} &= \overline{\mu_{H,(0,2)}} +  \overline{\mu_{H,(0,1)}}(1- \overline{\mu_{H,(0,1)}}) \\
        &= C_{02}\overline{\mu_{+,(0,2)}} +  C_{01}\overline{\mu_{+,(0,1)}}(1- C_{01 }\overline{\mu_{+,(0,1)}}) \\
        &= C_{02} \kappa_{+,(0,2)} + (C_{02} - C_{01}^2) \kappa_{+,(0,1)}^2 +  (C_{01}-C_{02}) \kappa_{+,(0,1)}  
    \end{split}
\end{align}

Just as in Eqs. (\ref{eq:depl-cond-cumu-start})-(\ref{eq:depl-cond-cumu-end}), the cumulants $\kappa_{h,(k,w)}$ follow as
\begin{align}
     \kappa_{h,(1,0)} &=  \kappa_{H,(1,0)} - \rho \langle c \rangle \kappa_{H,(0,1)} \\
    \kappa_{h,(0,1)} &=  (1-\rho) \kappa_{H,(0,1)} \\
        \kappa_{h,(2,0)}  &= (\rho(\rho-1)\langle c\rangle ^2+\sigma_c^2)\kappa_{H,(0,1)} + \rho^2 \langle c \rangle^2 \kappa_{H,(0,2)} -2\rho \langle c \rangle\kappa_{H,(1,1)} + \kappa_{H,(2,0)}\\
    \kappa_{h,(1,1)} &=  (1-\rho) (\kappa_{H,(1,1)} - \rho \langle c \rangle (\kappa_{H,(0,1)} + \kappa_{H,(0,2)}) ) \\
    \kappa_{h,(0,2)} &=  (1-\rho)^2 \kappa_{H,(0,2)} - \rho (1-\rho)\kappa_{H,(0,1)} 
\end{align}

Next, we use the cumulants $\kappa_{q|0,(k,w)}$ of the release size distribution (Eqs. \ref{eq:depl-gamma-size-mean-m} - \ref{eq:depl-gamma-size-var-v}) and the fact that $\kappa_{+,(k,w)} = \kappa_{q|0,(k,w)} + \kappa_{h,(k,w)}$, in order to obtain a system of equations for $\kappa_{+,(k,w)}$:
\begin{align}
    \kappa_{+,(1,0)} &=  V\rho \langle c\rangle + C_{10} \kappa_{+,(1,0)} - \rho \langle c \rangle C_{01} \kappa_{+,(0,1)} \\
    \kappa_{+,(0,1)} &=  V\rho + (1-\rho) C_{01} \kappa_{+,(0,1)} \\
    \begin{split}
        \kappa_{+,(2,0)}  &= V\rho(\sigma_c^2 + (1-\rho) \langle c \rangle^2) + (\rho(\rho-1)\langle c\rangle ^2+\sigma_c^2)C_{01} \kappa_{+,(0,1)} \\
        &\quad+ \rho^2 \langle c \rangle^2 ( C_{02} \kappa_{+,(0,2)} + (C_{02} - C_{01}^2) \kappa_{+,(0,1)}^2 +  (C_{01}-C_{02}) \kappa_{+,(0,1)}  ) \\
        &\quad-2\rho \langle c \rangle (C_{11}\kappa_{+,(1,1)} + (C_{11}- C_{10} C_{01}) \kappa_{+,(1,0)} \kappa_{+,(0,1)}) \\
        &\quad + C_{20} \kappa_{+,(2,0)} + (C_{20} - C_{10}^2) \kappa_{+,(1,0)}^2 +  (C_{10}-C_{20}) \kappa_{+,(1,0)}
    \end{split} \\
    \begin{split}
        \kappa_{+,(1,1)} &=  V(1-\rho)\rho \langle c\rangle +(1-\rho) (C_{11}\kappa_{+,(1,1)} + (C_{11}- C_{10} C_{01}) \kappa_{+,(1,0)} \kappa_{+,(0,1)} \\
        &\quad - \rho \langle c \rangle (C_{01} \kappa_{+,(0,1)} + C_{02} \kappa_{+,(0,2)} + (C_{02} - C_{01}^2) \kappa_{+,(0,1)}^2 +  (C_{01}-C_{02}) \kappa_{+,(0,1)} ) )
    \end{split} \\
    \kappa_{+,(0,2)} &=  V(1-\rho)\rho + (1-\rho)^2 (C_{02} \kappa_{+,(0,2)} + (C_{02} - C_{01}^2) \kappa_{+,(0,1)}^2 +  (C_{01}-C_{02}) \kappa_{+,(0,1)} ) - \rho (1-\rho)C_{01} \kappa_{+,(0,1)} 
\end{align}
This can be solved recursively:
\begin{align}
    \kappa_{+,(0,1)} &= \frac{V\rho}{1-(1-\rho)C_{01}} \\
    \mu_+:=\kappa_{+,(1,0)} &= \frac{1-C_{01}}{1-C_{10}} \langle c\rangle \kappa_{+,(0,1)} \\
    \kappa_{+,(0,2)} &= \frac{(1-\rho)}{1-(1-\rho)^2C_{02}}\Big(\rho(V-C_{01} \kappa_{+,(0,1)}) + (1-\rho) \Big( (C_{02} - C_{01}^2) \kappa_{+,(0,1)} ^2 +  (C_{01}-C_{02}) \kappa_{+,(0,1)} \Big)\Big) \\
    \begin{split}
        \kappa_{+, (1,1)} &= \frac{1-\rho}{1-(1-\rho)C_{11}} \Big(V\rho \langle c\rangle + (C_{11}- C_{10} C_{01}) \kappa_{+,(1,0)} \kappa_{+,(0,1)} \\
        &\quad - \rho \langle c \rangle (C_{02} \kappa_{+,(0,2)} + (C_{02} - C_{01}^2) \kappa_{+,(0,1)}^2 +  (2C_{01}-C_{02}) \kappa_{+,(0,1)} )  \Big)
    \end{split} \\
    \begin{split}
        \sigma_+^2:=\kappa_{+,(2,0)}  &= \frac{1}{1-C_{20}} \Big(V\rho(\sigma_c^2 + (1-\rho) \langle c \rangle^2) + (\rho(\rho-1)\langle c\rangle ^2+\sigma_c^2)C_{01} \kappa_{+,(0,1)} \\
        &\quad+ \rho^2 \langle c \rangle^2 ( C_{02} \kappa_{+,(0,2)} + (C_{02} - C_{01}^2) \kappa_{+,(0,1)}^2 +  (C_{01}-C_{02}) \kappa_{+,(0,1)}  ) \\
        &\quad-2\rho \langle c \rangle (C_{11}\kappa_{+,(1,1)} + (C_{11}- C_{10} C_{01}) \kappa_{+,(1,0)} \kappa_{+,(0,1)}) \\
        &\quad  + (C_{20} - C_{10}^2) \kappa_{+,(1,0)}^2 +  (C_{10}-C_{20}) \kappa_{+,(1,0)} \Big)
    \end{split} 
\end{align}

In the fixed-interval case $C_{kw} = e^{-(k+w\xi)/\lambda}$, this simplifies to:
\begin{align}
    \kappa_{+,(0,1)}^{(F)} &= \frac{V\rho}{1-(1-\rho)e^{-\xi / \lambda}} \\
    \mu_+^{(F)}=\kappa_{+,(1,0)}^{(F)} &= \frac{\langle c\rangle V\rho (1-e^{-\xi / \lambda})}{(1-e^{-1/\lambda})(1-(1-\rho)e^{-\xi / \lambda})} \label{eq:post-release-F-mean-new}\\
     \kappa_{+,(0,2)}^{(F)}  &= \frac{V\rho(1-\rho)(1-e^{-\xi/\lambda})}{(1-(1-\rho)e^{-\xi / \lambda})^2} \\
    \begin{split}
         \kappa_{+,(1,1)}^{(F)} &= \frac{\langle c \rangle V \rho(1-\rho) (1-e^{-\xi/\lambda})^2}{(1-(1-\rho)e^{-(\xi+1)/\lambda})(1-(1-\rho)e^{-\xi/\lambda})^2}
    \end{split} \\
    \begin{split}
         \sigma_+^{2(F)} =\kappa_{+,(2,0)}^{(F)}  &=  \frac{\mu_+^{(F)}}{1+e^{-1/\lambda}} \Big( e^{-1/\lambda} +  \frac{\langle c \rangle \left((1-\rho)(e^{-(1+2\xi)/\lambda}+1)+e^{-\xi/\lambda}(2\rho-1)  -e^{-(\xi+1)/\lambda}(1-\rho^2)\right)}{(1-(1-\rho)e^{-\xi/\lambda})(1-(1-\rho)e^{-(\xi+1)/\lambda})} \\
        &\quad + \frac{(1+ \rho e^{-\xi/\lambda}) \sigma_c^2}{ (1-e^{-\xi/\lambda})\langle c \rangle}\Big)
    \end{split} 
\end{align}

Matching the post-release mean in Eq.~(\ref{eq:post-release-F-mean-new}) to the previous result without vesicle depletion in Eq.~(\ref{eq:post-release-F-mean}), we find a modified mean release size of 
\begin{equation}
    \langle m^{(F)} (\lambda,\xi)\rangle = \langle c \rangle V \rho  \cdot \frac{1 - e^{-\xi / \lambda}}{1-(1-\rho)e^{-\xi /\lambda}}. \label{eq:mF-modified}
\end{equation}
Note, however, that this correction really only captures the post-release mean. Plugging Eq.~(\ref{eq:mF-modified}) into higher moment equations for the model without depletion yields results different from the ones obtained here.  

In the Poisson case, we specialize $C_{kw} = \frac{\lambda}{\lambda + k+w\xi}$ and find: 

\begin{align}
        \kappa_{+,(0,1)}^{(P)} &= \frac{V \rho(\lambda +\xi)}{\rho \lambda + \xi}\\
    \mu_+^{(P)}=\kappa_{+,(1,0)}^{(P)} &= \frac{\langle c\rangle V\rho\;\xi(\lambda+1)}{\rho\lambda+\xi} \label{eq:post-release-P-mean-new} \\
     \kappa_{+,(0,2)}^{(P)} &= \frac{V\rho(1-\rho)\xi \Big(2\xi(\lambda+\xi) + \rho\lambda(\lambda+\xi+(1-\rho)(\lambda+V\xi))\Big)}{(\rho\lambda(2-\rho)+2\xi)(\rho \lambda + \xi)^2} \\
     \kappa_{+,(1,1)}^{(P)} &= \frac{\langle c\rangle V \rho (1-\rho) \xi^2 \Big(V\rho\lambda(\rho(1-\rho)\lambda+(2-\rho)\xi- \rho)+ 2(1+\lambda+\xi)(\lambda\rho + \xi) \Big)}{(1+\rho\lambda +\xi)(\rho\lambda(2-\rho)+2\xi)(\rho \lambda+\xi)^2}\\
\begin{split}
 \sigma_+^{2(P)} = \kappa_{+,(2,0)}^{(P)}&= \mu_+^{(P)} \Big(\frac{\lambda  }{2  (\lambda +1)
} \\
&+ \frac{
V\rho \langle c\rangle\lambda \xi \left(
\rho \left(\rho + (1-\rho)^2 \lambda \right)
- (4 - \rho)\rho \xi
+ \xi(\xi+1)
\right)
}{
(\lambda +1)
\left(( 2 - \rho)\rho \lambda + 2 \xi\right)
(\rho \lambda + \xi)
(1 + \rho \lambda + \xi)
}\\
&+ \frac{
\langle c \rangle(\lambda +2)\left(
(2 - \rho)\rho \lambda (1 + \rho \lambda)
+ \left(2(1-\rho)+ \rho^2 \lambda)\right)\xi
+ 2 (1 - \rho)\xi^2
\right)
}{
2 (\lambda +1)
\left(( 2 - \rho)\rho \lambda + 2 \xi\right)
(1 + \rho \lambda + \xi)
} \\
&+ \frac{ \sigma_c^2 (\lambda +2)   (\lambda + \rho \lambda + \xi)}{2 \langle c \rangle (\lambda +1)
\xi  }\Big)
    \end{split} 
\end{align}

Again, we may compare Eq.~(\ref{eq:post-release-P-mean-new}) and (\ref{eq:post-release-P-mean}) and find a modified mean release size of
\begin{equation}
    \langle m^{(P)} (\lambda,\xi)\rangle = \langle c \rangle V \rho \cdot \frac{\xi}{\xi + \lambda \rho}.
\end{equation}

\section{\label{sec:appendix-sampling} Sampling of Stochastic Trajectories}

For efficient sampling of stochastic trajectories, we take advantage of three facts. First, there are far more degradation events (backward steps) than vesicle release events (forward steps). Second, the rate of forward steps does not depend on the number of backward steps taken. Third, the exact timing of the backward steps between two subsequent forward jumps does not affect the hitting probability and hitting time statistics. 

Thus, we can essentially separate the forward process and the backward process.
First, we determine the time of the next forward step by considering the probability $F(t, \tau) = \int_{0}^{\tau} f(t,s) ds$ for the next forward step occuring before time $t + \tau$. For a general Poisson with rate $\lambda(t)$, this is given as 
\begin{align}
    F(t, \tau) =1- \exp\left(  \int_{t}^{t+\tau} \lambda(s) \;ds \right).
\end{align}
As in the well-known Gillespie algorithm \cite{gillespie1977exact}, we set $F(t, \tau) = r$, where $r \sim \text{Unif}(0,1)$, and solve for $\tau$. 

Next, we determine the number of backward steps taken between $t$ and $t + \tau$ by drawing directly from the solution to the backward process $n(t+\tau) \sim\mathcal{B}^{n(t)}_{n(t+\tau)}(\tau)$, which is a binomial distribution with $N=n(t)$ and $p = e^{-\tau}$. If we account for vesicle depletion, we also update the number of empty release sites according to $u(t+\tau) \sim\mathcal{B}^{u(t)}_{u(t+\tau)}(\xi\tau)$. The post-release count is then $n(t+\tau) + m$, where $m$ is sampled from the release size distribution $q$, which may be conditioned on $u(t+\tau)$. The process of sampling forward steps is repeated, until either no more forward steps occur ($\tau = \infty$) or until an arbitrary steady state timeout is reached.

%\section{References}

%\bibliography{apssamp.bib}
